\documentclass[useAMS,usenatbib]{mn2e}
\usepackage{epsfig}
\usepackage{amsmath}    
\usepackage{amssymb}
\usepackage{natbib}
\usepackage{myaasmacros}
\def\Msol{\mbox{M$_\odot$}}

\title[Detecting high redshift quasars in X-rays]{Detecting quasars 
at very high redshift with next generation X-ray telescopes}
\author[K.J.  Rhook \& M.G.  Haehnelt]{Kirsty J.
  Rhook\thanks{krhook@ast.cam.ac.uk} \& Martin G.
  Haehnelt\thanks{haehnelt@ast.cam.ac.uk}\\Institute of Astronomy,
  Madingley Road, Cambridge CB3 0HA}
\begin{document}

\date{Revised Version May 13th, 2008}

\pagerange{\pageref{firstpage}--\pageref{lastpage}} \pubyear{2005}
  
\maketitle

\label{firstpage}

\begin{abstract}
  The next generation of X-ray telescopes have the potential to detect
  faint quasars at very high redshift and probe the early growth of
  massive black holes (BHs). We present modelling of the evolution of
  the optical and X-ray AGN luminosity function at $2<z<6$ based on a
  CDM merger-driven model for the triggering of nuclear activity
  combined with a variety of fading laws. We extrapolate the
  merger-driven models to $z>6$ for a range of BH growth scenarios. We
  predict significant numbers of sources at $z \sim 6$ with fluxes
  just an order of magnitude below the current detection limits and
  thus detectable with XEUS and Constellation-X, relatively
  independently of the fading law chosen.  The predicted number of
  sources at even higher redshift depends sensitively on the early
  growth history of BHs.  For passive evolution models in which BHs
  grow constantly at their Eddington limit, detectable BHs may be rare
  beyond $z \sim 10$ even with Generation-X. However, in the more
  probable scenario that BH growth at $z>6$ can be described by
  passive evolution with a small duty cycle, or by our merger driven
  accretion model, then we predict that XEUS and Generation-X will
  detect significant numbers of black holes out to $z \sim 10$ and
  perhaps beyond.
\end{abstract}

         
\section{Introduction}

The next generation of X-ray satellites promise to be powerful probes
of quasar activity out to very high redshifts. The European lead
project XEUS\footnote{http://www.rssd.esa.int/index.php?project=XEUS}
is designed to reach a sensitivity $\sim 100$ better than the
deepest observations to date. Constellation-X\footnote
{http://constellation.gsfc.nasa.gov/} would offer at least an order 
of magnitude improvement in sensitivity,
hopefully paving the way for a mission like Generation-X\footnote
{htt://www.cfa.harvard.edu/hea/genx.html} which could probe 50 times 
fainter again.  

In this paper we aim to assess the prospects for probing the early
growth of supermassive black holes (SMBHs) at $z \gtrsim 6$ with these
telescopes or similar future X-ray missions [see, e.g.,
\citet{haiman1999} and \citet{wyithe2003} for earlier work on
this]. The observed correlations between SMBH mass and galaxy
properties, which are now well established
\citep[e.g.][]{gebhardt2000, ferrarese2000}, strongly suggest a tight
link between the build-up of the stellar mass in galactic bulges and
the mass of the central BH. Most models of the growth history of SMBHs
assume that the frequent merging of galaxies predicted by CDM-like
hierarchical models of galaxy formation plays an important role in
this process \citep[e.g.][]{kh2000, wyithe2003, volonteri2003,
croton2006, malbon2006}. Such models, although subject to considerable
degeneracies, have been reasonably successful in reproducing the
observed luminosity function of AGN as well as the inferred BH mass
function and their respective evolution with redshift. These models
generally struggle, however, to combine the very efficient high
redshift growth of BHs, suggested by the SMBHs detected at $z \sim 6$,
with the strong feedback required to reproduce the rapid decline in
the fuelling rate of the most massive BHs at low redshift
\citep{bromley2004,malbon2006}.  Furthermore, at low redshift there is
considerable observational evidence that some, and maybe even most, of
the fuelling of SMBHs is not connected to major mergers between
galaxies \citep[e.g.][]{kauffmann2003,li2007}. This may be related to
the fact that merger-driven models of quasar activity have difficulty
reproducing the luminosity dependent density evolution (LDDE) observed
for faint AGN at redshifts below $z \sim 3$
\citep[e.g.][]{ueda2003,hasinger2005,silverman2007,bongiorno2007}
[see, e.g., \citet{marulli2007}], although more sophisticated models
which follow the simultaneous growth of galaxies and AGN can
incorporate feedback in such a way to orchestrate this effect
\citep{fontanot2006}.

We explore predictions for a variety of models for the early growth of
SMBHs which range from merger driven accretion to continuous
Eddington-limited accretion.  The BH growth scenario is crucial in
determining the number of active BHs that we expect to detect as faint
X-ray sources. However, as X-ray observations push to fainter flux
limits they are expected to discover a mixture of low-luminosity
objects at low to intermediate redshift, as well as distant objects
previously at undetectable redshifts. Predictions for the redshift
distribution of faint sources, which encrypt the growth history of
SMBHs, therefore require a simultaneous study of the faint end of the
luminosity function at all redshifts.

We have adopted a hybrid approach to modelling the redshift evolution
of the X-ray emission associated with the fuelling of SMBHs. At low
redshift ($z \lesssim 2$) we use the observed X-ray luminosity
functions, which are now well established down to very faint
luminosities. At intermediate redshifts the faint end of the X-ray
luminosity function is still subject to some uncertainty due to the
difficultly of assigning redshifts to faint sources \citep{aird2008}.
At $2<z<6$ we have therefore adopted a CDM merger-driven model for the
evolution of the emission from SMBHs, constrained by the available
optical and X-ray data.  The efficiency of BH formation, the quasar
lifetime and accretion rate, and the quasar spectral energy
distribution (SED) are all free parameters in such a model.  The
relevant assumptions should obviously be guided by our empirical and
physical understanding of quasars, but there is still much
flexibility, particularly near the observational limits in redshift
and luminosity.

Observationally, faint AGN are more likely to display signatures of
obscuration by gas and/or dust than brighter sources
\citep{ueda2003,simpson2005,treister2006,maiolino2007}. Most 
semi-analytic models of quasar activity do not model 
obscuration effects in sufficient detail to be able to explain these
trends. \citet{hopkins2005b,hopkins2005d,hopkins2006b,hopkins2006c}
have, however, recently  presented a model for the quasar luminosity
function (QLF) at $z<5$, which includes prescriptions for quasar
luminosity dependent fading and absorption, to successfully reconcile
the optical and X-ray QLFs.  We follow a very similar approach and
adopt many of the same assumptions. However, whereas
\citet{hopkins2006b} extract the quasar
formation rate from the constraints provided by the observed QLF, we
use a cosmological galaxy merger rate from the extended
Press-Schechter formalism \citep[e.g.][]{lacey1993}. This will allow
us to extend our model to the redshifts ($z>6$) we are predominantly
concerned with in a well-motivated way. Semi-analytic models are an
efficient way of exploring the wide range of physically plausible
evolutionary scenarios for the QLF. As we will discuss in some detail,
the faint end of the luminosity functions, and therefore the redshift
distribution of faint X-ray sources, is very sensitive to the assumed
time dependence of the fading of the emission in different wave-bands
during galaxy mergers.  We utilise the flexibility of our model to
study the constraints on this fading rate provided by the optical,
soft and hard X-ray QLFs, and the unresolved Cosmic X-ray background
(CXRB).

We then go on to explore several models for BH growth at $z>6$ which are
consistent with the observed data at $0<z<6$ and span a wide
range of possibilities for the rate of BH growth at high redshift.

Our paper is structured as follows. In Section 2 we briefly review the
constraints on the optical and X-ray QLFs. We describe our merger
driven model for $0<z<6$ in Section~3, and the high redshift
extensions in Section~4.3. In the main body of Section~4 we calibrate
our model against the measured optical and X-ray QLFs. In Section~5 we
examine the consistency of our models with the CXRB and observed ${\rm
logN}-{\rm logS}$ X-ray source distribution, and present our
prediction for the source density detectable with future missions,
before concluding in Section~6.

Throughout this paper we adopt a cosmological matter density $\Omega_m
= 0.27$, baryonic matter density $\Omega_b = 0.044$, cosmological
constant $\Omega_{\Lambda} = 0.73$, present day Hubble constant $H_o
\equiv  100h$~km~s$^{-1}$~Mpc$^{-1} = 71$km~s$^{-1}$~Mpc$^{-1}$, a mass
variance on scales of $8 h^{-1}$~Mpc $\sigma_8 = 0.84$ and a
scale invariant primordial power spectrum (slope $n=1$).

\section{Observational constraints}

We are primarily interested in exploring the constraints on the early
growth of SMBHs that may be provided by future X-ray missions.  In
order to do this in a conservative way we tie our model to a wide
range of observational constraints on the growth history of SMBHs at
low and intermediate redshift offered by the QLF in the optical, soft
and hard X-ray bands, as well as the CXRB.

The luminosity function of optically selected QSOs is well measured up
to $z \gtrsim 4$, with constraints on the bright end that reach up to
$z \sim 6$ \citep{croom2004,wolf2003,fan2003,fan2004,jiang2007}.  The
constraints on the X-ray QLF extend to $z \sim 4$ at best, but deep
X-ray surveys surveys reach to considerably fainter limiting
magnitudes than optical surveys.  Combined constraints of wide-field
and deep surveys result in a dynamic range of $\sim 5$ orders of
magnitude in luminosity at $z \sim 2$ compared to $\sim 3$ orders of
magnitude at optical wavelengths.  The low luminosities reached by
deep X-ray surveys has allowed \citet{shankar2007} to constrain the
faint-end slope of the optical QLF at $z \sim 6$ from the dearth of
X-ray detections above $z \sim 4 - 5$.

The density of optically bright quasars peaks at $z \sim 3$ with
quasar activity decreasing towards both lower and higher redshift
\citep[e.g.][]{boyle2000,wolf2003,croom2004,richards2006}. The optical
QLF is well fit by a double power-law with the break luminosity
evolving self-similarly with redshift, although there is evidence for
the bright end slope flattening toward high redshift
\citep{fan2004,richards2006} [but see also
\citet{fontanot2007a}]. More recent evidence that the space density of
lower luminosities sources peaks at $1<z<1.5$ \citep{bongiorno2007}
suggests that the full picture is probably more complicated, and that
the faint optically selected sources show a similar evolution to that
seen in X-ray surveys.  The X-ray luminosity function is best fit with
a luminosity dependent density evolution
\citep[e.g.][]{miyaji2001,ueda2003,fiore2003,hasinger2005,silverman2007}
with the source density peaking at $z \sim 0.7 - 1$. Additional
constraints at X-ray wavelengths come from the measurements of the
CXRB \citep{worsley2005,hickox2006}. Obviously the integrated emission
from faint AGN below the current detection limits cannot exceed the
measured unresolved CXRB and measurements of the background intensity
are thus probably the strongest constraints on the faint end of the
X-ray luminosity function.

\section{A hybrid model for  the evolution of the emission due to accretion
  onto supermassive black holes}

\subsection{Rationale for a hybrid model}

As motivated in the introduction, we have chosen a hybrid approach to
modelling the evolution of the emission due to accretion onto
SMBHs. At low redshift ($z<2$) we use the well established observed
X-ray luminosity functions, which extend to impressively faint
luminosities. At intermediate redshift, where the X-ray luminosity
function becomes less reliable, we use a merger-driven CDM-like model
with a range of assumptions for the decline of the accretion rate onto
the central BH during a merger. This model is calibrated using the
extensive optical data as well as the available X-ray data. In
Section~4 we describe an extension of the merger-driven model, as well
as models for growth via continuous Eddington limited accretion with a
range of duty cycles, to very high redshift ($z>6$).  Recently
\citet{shankar2007b} constructed an observationally anchored model for
the evolution of the SMBH population which indicates that the quasar
duty cycle increases with increasing redshift, providing some
empirical impetus for this approach.

We first discuss the ingredients of our merger-driven model in some
detail.

\subsection{A CDM-like merger driven model for  the luminosity  function  at  intermediate redshift}  

We assume that the accretion of gas onto SMBHs is predominantly
triggered by major galaxy mergers \citep{kh2000,dimatteo2005,
hopkins2005b}, and take the merger rate of dark matter haloes in the
standard $\Lambda CDM$ model for structure formation as a proxy for
the rate of galaxy mergers. This picture is admittedly simple but
has nevertheless been shown to yield a reasonably consistent model for
many of the properties of the observed QLF and its evolution at
intermediate redshifts \citep[e.g.][]{hnr1998,wyithe2002, wyithe2003,
volonteri2003,  croton2006, marulli2007}.

In a merger-driven model, new quasars continuously form at a redshift
and luminosity dependent rate. We assume that the quasars become
active with an initial peak luminosity $L_{\rm peak}$ and then fade
according to a ``fading law'', $\frac{dt}{dlogL}$.  For a given
model for the rate of formation of sources with peak luminosity
$L_{\rm peak}$, $\dot{n}(L_{\rm peak}) = \frac{d^2n}{dlogL_{\rm
peak}dt}$, the luminosity function, $\frac{dn}{dlogL}$, can be written
as
\begin{equation}\label{QLF_fading_law}
\frac{dn}{dlogL} = \int_{L}^{\infty}\frac{dt}{dlogL}(L,L_{\rm
  peak})\frac{d^2n}{dlogL_{\rm peak}dt}dlogL_{\rm peak}.
\end{equation}

We assume that $L_{\rm peak}$ is the Eddington luminosity of the final
mass BH. We further assume that dark matter halos host a central BH
with mass $M_{\rm bh}$ given by \citep[e.g.][]{wyithe2003}
\begin{eqnarray}\label{Mbh-Mhalo}\nonumber
M_{\rm bh} &=& \epsilon(M,z) M,\\
&=& \epsilon_o h^{\alpha/3} \left[ \frac{\Omega_m \Delta_c}{\Omega_m^z
    18 \pi^2} \right]^{\alpha/6}\\\nonumber
&&\times (1+z)^{\alpha/2} \left(\frac{M}{10^{12} M_{\sun}}\right)^{\alpha/3 - 1}M,\\\nonumber
\end{eqnarray}

\noindent where $\Omega_m^z=\frac{\Omega_m(1+z)^3}{\Omega_m(1+z)^3+
\Omega_\Lambda+\Omega_k(1+z)^2}$, $d \equiv \Omega_m^z-1$ and
$\Delta_c=18\pi^2+82d-39d^2$ is the overdensity of a virialised halo
at redshift $z$. This relationship is motivated by the observed
correlation between $M_{\rm bh}$ and the velocity dispersion of the
host galaxie's bulge, $\sigma$ \citep{ferrarese2002,
shields2003}. We assume that $\sigma$ may be approximated by $v_{\rm
vir}/\sqrt{2}$, where $v_{\rm vir}$ is the virial velocity of the dark
matter halo from \citet{bl2001}. Empirical
estimates of $\alpha$ typically fall in the range $4-5$. We have
chosen $\alpha = 5$ which is consistent with a simple self-regulated
growth scenario in which the BH grows until it radiates enough
energy to unbind the gas that is feeding it
\citep[e.g.][]{silk1998,hnr1998,wyithe2003}.
The merger  rate  $\frac{d^2n}{dlogL_{\rm peak}dt}$ is taken as the merger rate of
halos with final BH mass such that $L_{\rm peak} = L_{\rm Edd}(M_{\rm
bh})$. Using equation (\ref{Mbh-Mhalo}) to relate BH and host halo mass,
we may write

\begin{equation}\label{bol_QLF}
\frac{d^2n}{dlogL_{\rm peak}dt} = \frac{3}{\alpha} M 
\int_{\Delta M = 0.25(M-\Delta M)}^{0.5(M-\Delta M)}\dot{N}(M, \Delta M,
t)d\Delta M, 
\end{equation}

\noindent where the limits of integration ensure that we count only
major mergers and $\dot{N}(M, \Delta M, t)$ is the merger rate of dark
matter haloes of mass $M-\Delta M$ and $\Delta M$ per unit cosmic time
$t$,

\begin{equation}
N(M,\Delta M,t) = \left.\frac{d^2P}{d \Delta M dt} \right|_{M - \Delta M}\frac{dn}{d(M- \Delta M)}.
\end{equation}

\noindent Here $\left.\frac{d^2P}{d \Delta M dt} \right|_{M - \Delta
M}$ is the probability per unit time that a halo mass $\Delta M$ will
merge with another halo to form a halo with mass $M$ from
\citet{lacey1993}, and $\frac{dn}{d(M- \Delta M)}$ is the space
density of halos with the appropriate mass difference from
\citet{ps1974} (with modification from \citet{sheth1999}, see
\citet{rhook2006} and \citet{wyithe2003} for similar calculations).

Equation~(\ref{bol_QLF}) becomes inaccurate as $\frac{dt}{dlogL}$
approaches the Hubble time at the relevant redshift.  This could in
principle be relevant at high redshift, however at high redshift we
are mainly concerned with the bright end of the QLF for which the
fading time-scale remains short (see Section \ref{fading_laws}).
\noindent Note that for $\frac{dt}{dlogL}(L,L_{\rm peak}) = {\rm constant}$
(corresponding to an exponential light-curve) or
$\frac{dt}{dlogL}(L,L_{\rm peak}) \propto \delta(L-L_{\rm Lpeak})$
(corresponding to a top-hat light-curve) the shape of the QLF is
identical to that of $\dot{n}(L_{\rm peak})$.

\subsection{Gas accretion, quasar lifetime and fading laws, and the
minimum black hole mass}\label{fading_laws}

As discussed in the previous section, the faint end of the QLF is very
sensitive to the assumed fading law. When two gas-rich galaxies merge,
the resulting tidal torques drive large amounts of gas into the
central region, providing fuel for the rapid growth of a SMBH.  The
fading of the quasar emission due the gas accretion onto a central
SMBH is governed by the infall of gas to the centre and its subsequent
accretion onto the SMBH during the late stages of the merger.
\citet{hopkins2005b} used numerical simulations of galaxy mergers,
with a prescription for the subsequent accretion of gas onto a central
SMBH from \citet{dimatteo2005}, to obtain a physically motivated
fading law.  \citet{hopkins2005b} present a power-law representation
for this fading law, $\frac{dt}{dlogL}$, which has been fit to the
results of several hundred simulations of mergers between equal mass
galaxies resulting in quasars with peak bolometric luminosities
between $10^{8}$ and $10^{15}$~L$_{\sun}$. The power-law index depends
on $L_{\rm peak}$, with higher luminosity quasars expected to spend
relatively more time close to their Eddington limit,

\begin{eqnarray}\label{hopkins_fading}
\frac{dt(L,L_{\rm peak})}{dlogL} &=& \left| \alpha_L 
 \right|t_9\left( \frac{L}{10^9~{\rm L}_{\odot}}\right)^{\alpha_L(L_{\rm peak})},\\ 
\alpha_L(L_{\rm peak}) &=& -0.95 + 0.32{\rm log}_{10}(L_{\rm
  peak}/10^{12}~{\rm L}_{\odot}),
\\\nonumber
\end{eqnarray}

\noindent where $t_9 \approx 10^9~{\rm yrs}$ and $\alpha_L \geq -0.2$. 
\citet{hopkins2005b} model the quasar luminosity over the entire
merger process and therefore equation~(\ref{hopkins_fading}) includes
the activity of the quasar prior to the luminosity reaching $L_{\rm
peak}$. This distinction should not be important for our modelling.

In order to investigate the effect of the quasar light-curve on the
the faint end of the luminosity function, we consider the fading law
as proposed by  \citet{hopkins2005b} as well  as a fading law where 
quasars fade more rapidly and spend  equal amounts of time at each 
luminosity (relative to the peak luminosity) independently of $L_{\rm peak}$,

\begin{equation}\label{constL_fading}
\frac{dt}{dlogL} = \left| \alpha_L \right| t_{q} \left( \frac{L}{L_{\rm peak}} \right)^{\alpha_L}.
\end{equation}

\noindent Here $\alpha_L$ is a non-positive constant and larger values
of $|\alpha_L|$ result in light curves for which quasars spend
relatively more time at sub-Eddington luminosities. Note that values
of $\alpha_L$ that are much less than zero are excluded by the
observational data.

\begin{figure*}\label{plot1}
  \vspace*{70mm} \includegraphics{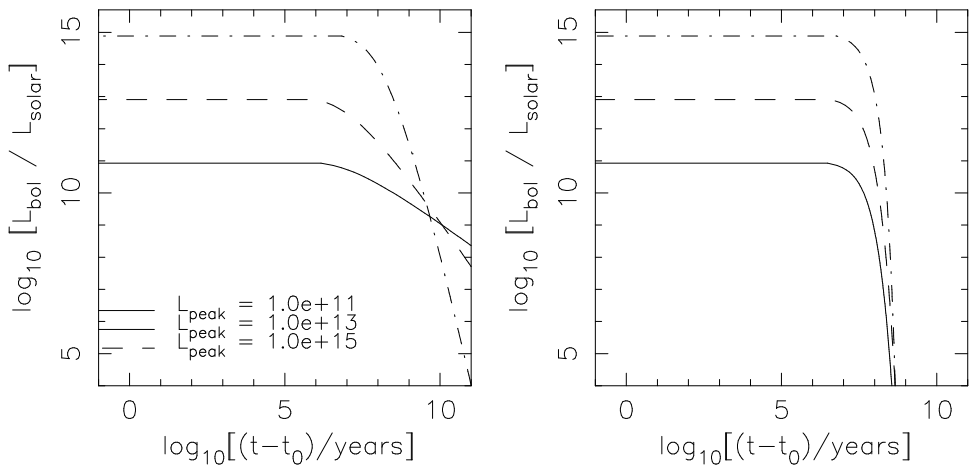}
\caption{Light curves used to generate model QLFs for quasars with intrinsic,
peak, bolometric luminosities of $10^{11}$ (\emph{solid} line),
$10^{13}$ (\emph{dashed} line) and $10^{15}$ (\emph{dot-dashed} line)
$L_{\odot}$. The left panel shows the slow  fading law,
corresponding to equation~(\ref{hopkins_fading}) with the average lifetime ($t_9$)
increased by a factor of two. The right panel is for the rapid fading law
$\frac{dt}{dlogL} = 1.7 \times 10^{7}~{\rm yrs} \left(\frac{L}{L_{\rm
peak}}\right)^{-0.01}$ corresponding to rapid (almost exponential) luminosity independent
fading.}
\end{figure*}

The faint end of the luminosity function is also sensitive to the
minimum mass of a BH powering a quasar, $M_{\rm bh,min}$, which for
Eddington limited growth is equivalent to the minimum peak luminosity
of a quasar, $L_{\rm peak,min}$. There are very few claimed detections
of SMBHs with masses smaller than $10^{6}\Msol$ \citep[but
see][]{greene2007}, and \citet{greene2007a} find that the mass
function of local low-luminosity active BHs turns over at $M_{\rm bh}
\sim 10^{6}$~$\Msol$. It is also a matter of intense debate whether
the build-up of SMBHs by hierarchical merging extends to intermediate
mass BHs in the mass range $100~\Msol - 10^{6}~\Msol$
\citep[e.g.][]{haehnelt2004,volonteri2007}. We therefore also explore
the effect of varying the minimum BH mass, $M_{\rm bh,min}$, in our
models.

\subsection{Modelling the Spectral Energy Distribution} \label{SED}

To relate the intrinsic bolometric QLF to observational data, we also
require a model for the intrinsic SED and a prescription for the
relative obscuration of quasars at different wavelengths.

The analyses of samples of optically selected (type-I) quasars suggest
that there is a luminosity dependent correlation between the X-ray and
optical luminosity, which may be described in terms of the power-law
index connecting the optical and X-ray flux,
$\alpha_{OX}$. $\alpha_{OX}$ may be written
\citep[e.g.][]{steffen2006}

\begin{eqnarray}\label{LB-LX}\nonumber
\alpha_{OX} &=& - \frac{{\rm log}_{10}(L_{\nu}(2500~{\rm \AA})/L_{\nu}(2~{\rm keV}))}{{\rm log}_{10}(\nu(2500~{\rm \AA})/\nu(2~{\rm keV}))},\\
&=&- A {\rm log}_{\rm 10}\left(\frac{L_{\nu}(2500~{\rm \AA})}{{\rm ergs/s/Hz}}\right) + B,
\end{eqnarray}

\noindent where $A, B > 0$ corresponding to an anti-correlation of 
$\alpha_{OX}$ with optical flux.  \citet{steffen2006} estimate $A =
0.137 \pm 0.008$ and $B = 2.638 \pm 0.24$ for an optically selected
sample of quasars with no apparent intrinsic absorption or radio
activity. \citet{steffen2006} find no strong evidence for evolution of this
relationship with redshift, however the data is consistent with
$\alpha_{OX}$ decreasing slowly with redshift as $\sim 0.01z$.

We adopt a modified version of the model SED presented in
\citet*{hopkins2006c} to calculate the (luminosity dependent) SED for
a given observed frequency, $\nu_{\rm obs}$, quasar redshift, $z$, and
bolometric luminosity, $L_{\rm bol}$. This model assumes the  SED to 
be essentially a broken power-law which
mimics reprocessed emission in the NIR, the UV bump and the hard X-ray
excess features with an exponential cut-off at 500~keV.  The SED is
normalised so that the observed relation between optical and X-ray
emission described by equation~(\ref{LB-LX}) with $A =0.109$ and $B =
1.739$ is preserved.
  
Note that \citet{hopkins2006c} found that the luminosity dependence of
the relation between optical and X-ray emission is more important for
reconciling the optical and X-ray QLFs at $0<z<5$ than the detailed
modelling of the features in the SED.  The anti-correlation of
$\alpha_{OX}$ with $L_{\nu}(2500\AA)$ results in a narrower spread of
X-ray luminosities compared to the corresponding optical range. The
model X-ray QLF is thus steeper than the equivalent bolometric QLF.

  We find that the bright end of our model X-ray luminosity function
  becomes too steep compared to the measurements in both the soft and
  hard X-ray bands when we use the \citet{hopkins2006c} SED.
  Including a dispersion in the bolometric corrections, which
  \citet{hopkins2006c} note has the largest effect on the bright end
  of the X-ray QLF, would smooth the QLF and somewhat alleviate this.
  We have chosen instead to alter the luminosity dependence of the SED
  model to fit the data as described below.
  
  To achieve a good match with the bright end of the observed X-ray
  QLF we adopt an SED for $L_{\rm bol} > 10^{14}$~L$_{\sun}$ with $A =
  0.104$ and $B = 1.739$.  We use the same power-law indices,
  break-points and exponential cut-off as \citet{hopkins2006c} but
  have not modelled the reflection component or the NIR emission.
  This choice of parameters increases the power in the soft and hard
  X-ray bands by approximately 40~per~cent at $L_{\rm bol} =
  10^{14}$~L$_{\sun}$, increasing to 45~per~cent at
  $10^{15}$~L$_{\sun}$.  The  power at UV wavelengths changes
  minimally.  To ensure a smooth QLF
  we take an average of the \citet{hopkins2006c} and the above
  modification for bolometric luminosities between $10^{12}$ and
  $10^{14}$~L$_{\sun}$ weighted by the logarithmic bolometric
  luminosity. 

\subsection{The effect of absorption on the SED} \label{absorption}

Light emitted from the accretion disk of the AGN is reprocessed during
transmission through its host galaxy, the IGM and our own galaxy. At
the X-ray energies and redshifts we are mainly interested in (observed
frame $0.5 - 10$~keV) the absorption of X-rays by neutral hydrogen in
the IGM can be neglected. Galactic absorption is generally corrected
for, and therefore we are mainly concerned with absorption at the
quasar redshift. Assuming that emission at all wavelengths is obscured
by the same body of gas, the absorption at a particular wavelength can
be determined given a column density distribution of neutral hydrogen
(which may depend on parameters of the system) a dust-to-gas ratio and
a reddening law for the dust absorption.  We again adopt similar
assumptions as \citet{hopkins2005d} for our model, which we discuss in
the following two sections.

\subsubsection{Column density distribution of the absorbing gas}

The column density distribution of absorbing gas is expected to be
related to the fading law; gas both feeds the BH and obscures the
radiation released. However the astrophysical relationship is
complicated and depends on, for example, the surrounding star
formation and the geometry of the accretion disc. \citet{hopkins2005b}
have used the numerical simulations of quasars fuelled by gas
accretion by \citet{dimatteo2005} to construct a parametrised
luminosity and time dependent model for the obscuring column
density. The probability $P$ that a quasar is obscured by a given
total hydrogen column density $N_{\rm H}$ depends on the amount of
time a quasar remains in a given accretion phase. The simulated data
is fit with a log-normal distribution

\begin{eqnarray}\nonumber\label{NH_dist}
P(N_{\rm H}, L) &=& \frac{1}{\sigma_{\rm N_{\rm H}}\sqrt{2\pi}} exp \left\{ -[{\rm log}_{10}(N_{\rm H}/\bar{N}_{\rm H})]^2/ \left(2\sigma_{N_{\rm H}}^2 \right)\right\},\\ \nonumber
\bar{N}_{\rm H} &=& 10^{21.9}~{\rm cm}^{-2}\left(\frac{L}{10^{11}~{\rm L}_{\odot}}\right)^{0.43},\\ 
\sigma_{N_{\rm H}} &\approx& 1.2.\\\nonumber
\end{eqnarray}

 \noindent Note that the mean column density, $\bar{N}_{\rm H}$, and
dispersion, $\sigma_{\rm N_{\rm H}}$, are predominantly determined by
the instantaneous bolometric luminosity of the quasar $L$.

\citet{hopkins2005b} also demonstrated that the shape of this
distribution for quasars with a B-band luminosity above
$10^{11}$~L$_{\sun}$ is consistent with a statistical analysis of the
reddening found toward SDSS quasars in \citet{hopkins2004}.  This
analytic fit is derived from simulations of gaseous disc galaxies, and
of course may not be representative of the typical accretion rate onto
black holes at all redshifts. However at the redshifts that we are
applying it ($z>2$) galaxies are expected to be gas rich.

\subsubsection{Dust and gas absorption }

Since we are predominantly comparing our model to optical data quoted
in UV magnitudes we have used the reddening law presented in
\citet{gaskell2004} to model dust and gas absorption. The reddening
law is normalised such that the optical depth in the V-band
($5500$~\AA) is the same as for an SMC-like reddening curve [taken
from \citep{pei1992}] for a galaxy with the metallicity of the Milky
Way. This results in a V-band optical depth identical to that adopted
by \citet{hopkins2005d} who assumed a SMC-like reddening curve with a
gas-to-dust ratio as for the Milky Way.  At X-ray wavelengths the
absorption is dominated by photo-electric absorption and Compton
scattering by hydrogen.  Following \citet{hopkins2005d} we assume an
average neutral gas fraction of $0.35$ and assume that the ionised
component contributes to the photoelectric absorption, but not the
optical reddening or Compton scattering. We note that this model
makes the simplistic assumption that the distribution of  gas and dust 
is spatially uniform.

\subsection{The model luminosity function taking absorption into account}\label{abs_LF}

We now turn to calculating the observed luminosity function for a
given redshift $z$ and observed wavelength $\nu$ (or more usefully for
comparison with X-ray observations, wave-band $\nu_a \rightarrow
\nu_b$).

Given the probability that a quasar with a given intrinsic bolometric
luminosity $L^{\prime}$ is obscured by a given column density, the
probability that the intrinsic specific luminosity $L^{\prime}_{\nu}$
is observed to have a specific luminosity $L_{\nu}$,
$P(L_{\nu}|L^{\prime}_{\nu})d{\rm log}L^{\prime}_{\nu}$, may be
written

\begin{eqnarray}\label{NH_distribution}\nonumber
&&P(L_{\nu}|L^{\prime}_{\nu})d{\rm log}L^{\prime}_{\nu}\\ \nonumber
 &=& P(N_{\rm H} = \frac{1}{\sigma_{\nu}}{\rm log}\left( \frac{L^{\prime}_{\nu}}{L_{\nu}}\right), L^{\prime}) \frac{d{\rm log}_{10} N_{\rm H}}{d{\rm log} L^{\prime}_{\nu}} d{\rm log}L^{\prime}_{\nu}, \\ \nonumber
&=& P(N_{\rm H} = \frac{1}{\sigma_{\nu}}{\rm log}\left(\frac{L^{\prime}_{\nu}}{L_{\nu}}\right), L^{\prime})\frac{{\rm log}_{10}(e)}{N_{\rm H}\sigma_{\nu}}d{\rm log}L^{\prime}_{\nu},
\end{eqnarray}

\noindent where $\sigma_{\nu}$ is the absorption cross section at
$\nu$ in units cm$^2$. The absorbed luminosity function at an
observed wavelength $\nu$ and redshift $z$,
$\left.\frac{dn}{dlogL_{\nu}}\right|_{obs}(z)$, can be calculated by
integrating over the unabsorbed bolometric QLF,
$\frac{dn}{dlogL_{\nu}}$,

\begin{eqnarray}\label{abs_lf}\nonumber
\left.\frac{dn}{dlogL_{\nu}}\right|_{obs}(z)&=& \int_{{\rm log}(L^{\prime}_{\nu(1+z)}) = {\rm log}(L_{\nu(1+z)})}^{\infty} \frac{dn}{dlogL^{\prime}_{\nu(1+z)}}(z)\\ 
&&P(L_{\nu(1+z)}|L^{\prime}_{\nu(1+z)}) d{\rm log}L^{\prime}_{\nu(1+z)}. 
\end{eqnarray}

The convolution is identical if we want to consider the luminosity
function in band $(\nu_a,\nu_b)$, except the change of variable in
equation~(\ref{NH_distribution}) becomes
$\frac{dlog_{10}N_H}{dlogL^{\prime}_{\rm band}} = \frac{1}{N_H
  <\sigma_{\nu}>}$. Here $<\sigma_{\nu}>$ is the average absorption
cross section in the band, weighted by the observed luminosity $\nu
L_{\nu} e^{-\sigma_{\nu}N_H}$,

\begin{equation}
<\sigma_{\nu}> = \frac{\int_{\nu_a}^{\nu_b}d \nu \sigma_{\nu} \nu L_{\nu}
e^{-\sigma_{\nu}N_H}}{\int_{\nu_a}^{\nu_b}d \nu \nu L_{\nu}
e^{-\sigma_{\nu}N_H}}.
\end{equation}

\subsection{The blowout phase}

The probability distribution of the absorbing hydrogen column density 
given in equation~(\ref{NH_dist}) will not be an adequate description
when the luminosity is close to  the peak luminosity, 
as much of the gas is expected to be blown away by radiation
pressure. As discussed by \citet{hopkins2005b} the true column density 
distribution in the simulations is bi-modal, with
quasars in the last e-folding of growth having a much lower obscuring
column density distribution. This so called ``blow-out'' phase lasts 
approximately 10 per cent of the total time the quasar spends
accreting. \citet{hopkins2005b} suggest that this bi-modality can
explain why optically selected quasars are observed to have lower
obscuring column densities for brighter quasars
\citep[e.g.][]{ueda2003}, counter to the positive correlation between
$L$ and $\bar{N}_H$ in equation~(\ref{NH_dist}).

We have taken a simple approach to modelling such a blowout phase; 
we assume that a fraction $f_{\rm blowout} = 0.1$ of quasars within 
a factor of $e$ of their peak luminosity
experience no intrinsic absorption. The resulting absorbed luminosity
function is then larger than that in equation (\ref{abs_lf}) by a term
$\frac{dn}{dlogL_{\nu}}(\frac{L}{L_{\rm peak}}> \frac{1}{e}) \times
f_{\rm blowout}$, where $\frac{dn}{dlogL_{\nu}}(\frac{L}{L_{\rm
    peak}>} \frac{1}{e})$ is the QLF of sources with specific
luminosity $L_{\nu}$ that have total bolometric luminosity within $e$
of their peak luminosity. $\frac{dn}{dlogL_{\nu}}(\frac{L}{L_{\rm
    peak}>\frac{1}{e}})$, increases with $L_{\rm bol}$, and approaches
the original (without blow out phase) value of
$\frac{dn}{dlogL_{\nu}}$ for large $L_{\rm bol}$. This
prescription for the blowout phase essentially puts a lower limit on
the ratio of the absorbed to unabsorbed luminosity function at high
luminosities equal to $f_{\rm blowout}$.

The parameters of the blow-out phase affect the relative number of
bright optical and X-ray luminous sources. In particular, if we didn't
include a blow-out phase (and were therefore assuming that more of the
bright optical sources are obscured) then we would predict more X-ray
bright sources.  However we find that the absorption in the soft X-ray
band behaves similarly to that in the optical and therefore if we did
not include the blow-out phase it would be very difficult to reproduce
the space density of quasars that are bright in soft X-rays.

\section{ Evolution of the  model  luminosity function in different
  wavebands and   comparison with observational data}
\begin{figure*}\label{plot2}
  \vspace*{120mm} \includegraphics{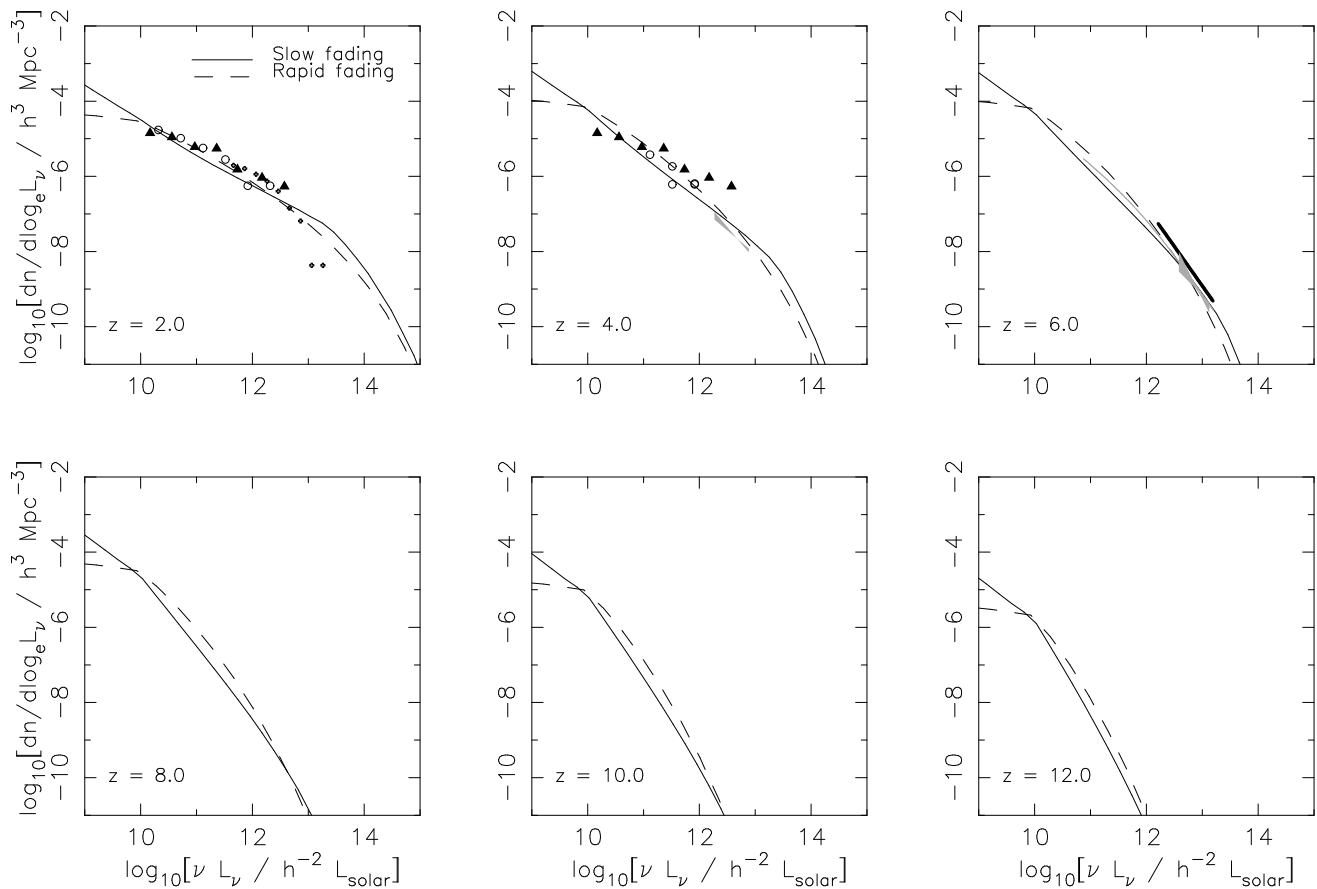}
\caption{Model rest-frame $1450$~\AA~QLF at $z = 2,4,6,8,10,12$ for a
merger driven model with a slow fading law (\emph{solid line}) and the
rapid fading law (\emph{dashed line}) as described in the text.  Note
that we have neglected absorption by neutral gas in the IGM which will
affect what is observed in practice at $z\gtrsim 6$. The hollow
circles are the measurement of the faint end QLF from the COMBO-17
survey (Wolf et al.~2003). The black triangles at $z=2,4$ show the
luminosity function measured by \citet{bongiorno2007} in the redshift
interval $2.1<z<3.6$. The small black diamonds at $z = 2$ are the
results from \citet{croom2004} using the SED from \citet{hopkins2006c}
to convert from B-band to rest frame 1450~\AA~ luminosities. The grey
bow-ties represent the constraints on the bright end QLF measured in
the SDSS survey (Fan et al. 2002, 2004). The thick black line at $z =
6$ is the fit from \citet{jiang2007}. The thin grey line at $z=6$ is
the Shankur \& Mathur~(2006) constraint on the slope of the QLF from
the dearth of X-ray sources.}
\end{figure*}

\begin{figure*}\label{plot3}
  \vspace*{120mm} \includegraphics{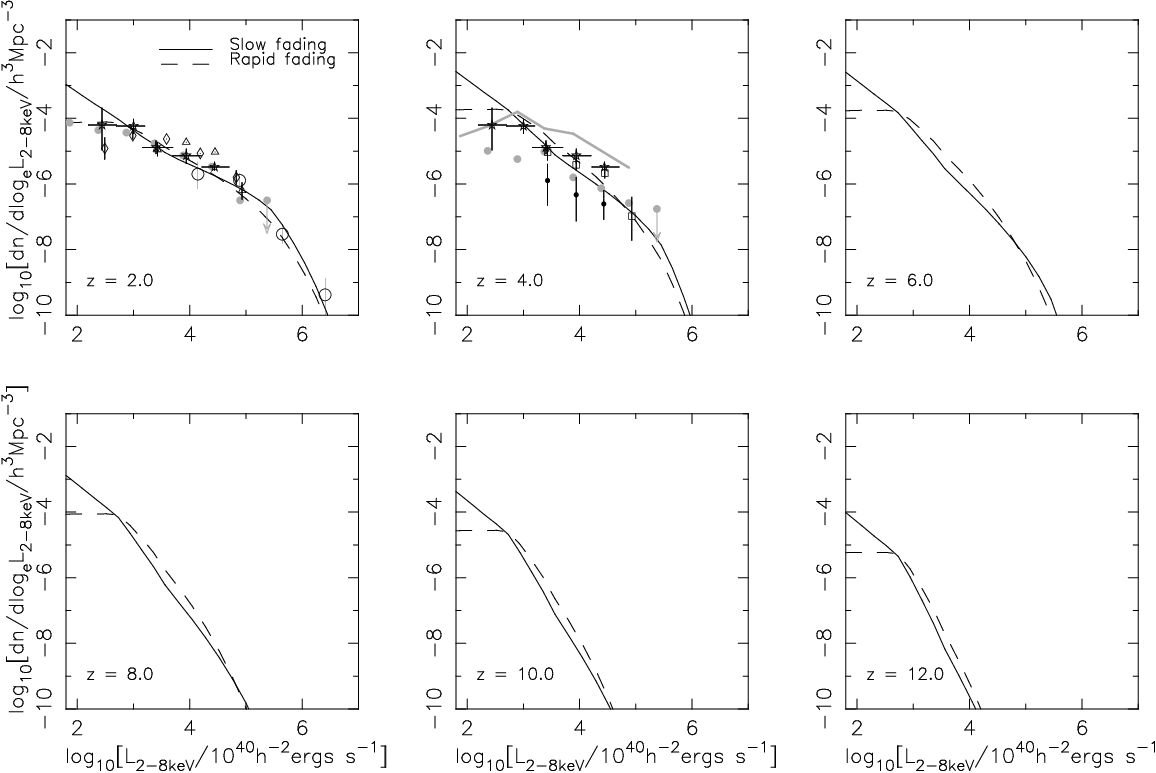}
\caption{Model rest-frame $2-8$~keV QLF at $z = 2,4,6,8,10,12$ for a
merger driven model with slow fading (\emph{solid line}) and rapid
fading (\emph{dashed line}) as in Figure~2. The filled grey circles
are estimates from \citet{barger2005} and the hollow grey circles the
rest frame $2-10$~keV data from \citet{ueda2003}.  The thick grey line
in the $z=4$ panel corresponds the estimate from \citet{barger2005}
assuming that all spectroscopically unconfirmed sources lie in the $z
= 3 - 5$ bin. The black markers are the hard X-ray QLF from
\citet{silverman2007}; the hollow diamonds for $1.5<z<2$, the hollow
triangles for $2<z<3$, the hollow squares for $3<z<4$ and the filled
circles for $4<z<5.5$. The stars with horizontal and vertical error
bars are the measurement for $2.5<z<3.5$ from \citet{aird2008}.}
\end{figure*}

\begin{figure*}\label{plot4}
  \vspace*{120mm} \includegraphics{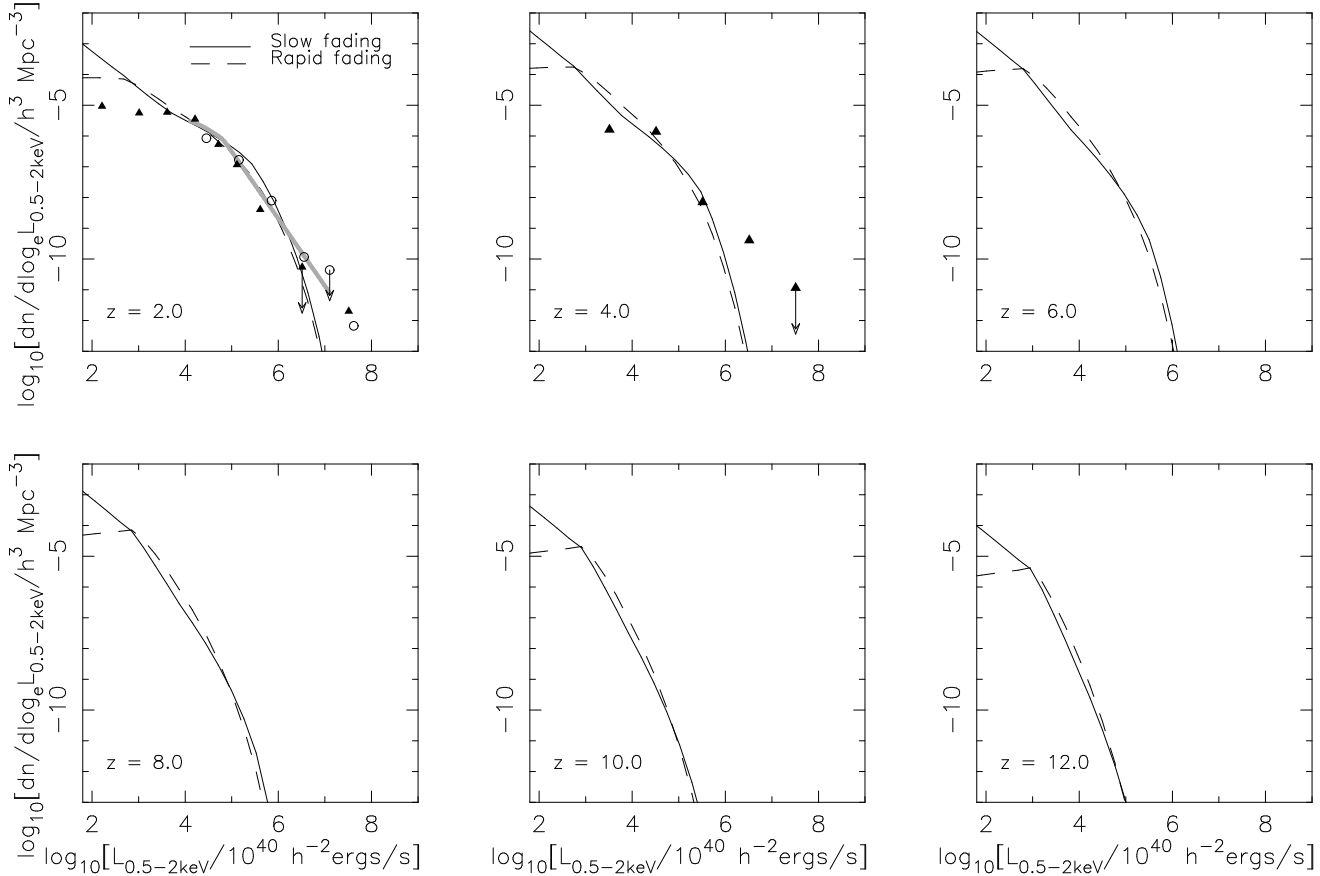}
\caption{Observed frame $0.5-2$~keV QLF at $z = 2,4,6,8,10,12$ for the
same models as in Figure~3. The thick grey line is the best-fit
luminosity dependent density evolution model from \citep{miyaji2000}
(LDDE2). The hollow markers are the binned luminosity function for $z
= 1.6 - 2.3$ from \citet{miyaji2001}. The black triangles denote the
type-I AGN QLF from \citep{hasinger2005}.}
\end{figure*}

\subsection{Calibrating the merger-driven model for the evolution
  of the luminosity function at intermediate redshift with observations}

We vary the parameters $(t_q, \epsilon_o, L_{\rm peak, min})$ in order
to obtain an acceptable fit to the observed luminosity functions at
rest-frame $1450$~\AA, and in the soft and hard X-ray bands. We
determine the normalisation of our models (governed by $t_q$ and
$\epsilon_o$) by comparing to the observed optical QLF, as these
constraints span the broadest redshift range.  However, as we will
discuss below, the optical data has little power to constrain $L_{\rm
peak,min}$ and constraints on this value come from  low-luminosity X-ray
observations alone.

We find that assuming the luminosity independent fading law in
equation~(\ref{constL_fading}), it is difficult to accommodate
values of $\alpha_L$ much less than zero, due to the intrinsic
steepness of the cosmological merger rate we are adopting for the
quasar formation rate. However for $\alpha_L = -0.01$, $t_q = 1.74
\times 10^{7}$~years and $\epsilon_0 = 10^{-5.05}$ we recover a good
fit to the optical QLF at $z = 2 - 6$ (see Figure 2). This value of
$\alpha_L$ corresponds to a light-curve for which the luminosity drops
off almost exponentially with time, we will therefore refer to this
model as the ``rapid fading'' model. The corresponding light curve is
compared to the \citet{hopkins2005b} light curve (with some adjustment
described below) in Figure 1. The \citet{hopkins2005b} fading law,
with appropriate $\epsilon_o$, also provides a reasonable fit to the
optical data at $z = 2-6$ as described by
equation~(\ref{hopkins_fading}). To put it on equal footing with what
we have done for the fit with the rapid fading law we allow the
characteristic time, $t_9$, to vary. We find that the fit is improved
when $t_9$ is increased by a factor of 2 to $2\times 10^9$~years. A
slightly larger value of $\epsilon_o = 10^{-4.97}$ than for the rapid
fading model is required to offset the smaller amount of time that
bright quasars spend at their Eddington luminosity. We refer to the
model with this fading law as the ``slow fading'' model. Since the two
models provide comparable fits to the optical data, the optical data
alone appears to offer little power to constrain the luminosity
dependence of the fading law.

Note that there is an excess in the predicted number of optically
bright quasars at $z = 2$ for each fading law. The most plausible
explanation for this discrepancy between our model and the data is
probably our neglect of AGN feedback. AGN inject large amounts of heat
into their surroundings \citep[e.g.][]{dunn2006,best2007} and should
be capable of suppressing cooling flows in dark matter halos with
masses above $\sim {\rm few} \times 10^{13}h^{-1}$~M$_{\sun}$ -
resulting in the formation of groups and clusters of galaxies rather
than super-sized quasars in very large dark matter halos \citep[see,
e.g.][]{sijacki2006, sijacki2007, rines2007}.  This effect is expected
to be more prominent at low redshifts when the typical masses of
merging dark matter halos is largest. Indeed most semi-analytic models
of the quasar population require some arbitrary high mass cut-off in
order to fit to the low redshift data \citep[e.g.][]{kh2000,
cattaneo2007}. To capture this effect in our model in a redshift
dependent way would be difficult and we do not attempt this here. It
is also for this reason that our model, and similar semi-analytic
models for the QLF, do not reproduce quasar luminosity function below
$z \sim 2$ very well.

With $t_q$ and $\epsilon_o$ chosen to reproduce the optical luminosity
functions we find fairly good agreement with the observed $z \sim 2$
QLF in the hard X-ray band, and consistency with the constraints on
the $z \sim 4$ hard X-ray QLF (see Figure 3).  The value of the
minimum BH mass (or minimum peak luminosity) does not affect the QLF
at the luminosities relevant for optical constraints, however the
constraints on the X-ray QLF are approximately 2 orders of magnitude
deeper at $z = 2$ than in the optical. We find that a cut-off at
$L_{\rm peak} = 10^{11}$~$L_{\sun}$ is required for the rapid fading
model in order to avoid over-predicting the number of sources in the
lowest luminosity bin in the hard X-ray QLF at $z = 2$.  We apply the
same cut-off in $L_{\rm peak}$ to the Hopkins-fading model and find it
over-predicts the rather shallow faint end of the observed luminosity
function at $z = 2$ in the hard X-ray band (see Figure 3). Note that
an Eddington luminosity of $10^{11}$~L$_{\sun}$ corresponds to a BH
mass $\sim 3.10^6$~M$_{\sun}$ and there is certainly evidence for the
existence of BHs with masses smaller than this
\citep[e.g.][]{greene2007}.

In Figure~4 we plot our model observed frame $0.5 - 2$~keV QLF and
compare it to the results of \citet{miyaji2000}, \citet{miyaji2001}
and \citet{hasinger2005}. Even with the increased  power at X-ray
wavelengths for the brightest sources, we still seem to under-predict
the space density in the brightest bins at $z = 2$ and $4$. Note,
however that the highest L data point is at each redshift is
calculated from only once source with an extremely high soft X-ray
luminosity of $\sim 10^{14}$~L$_{\sun}$, and therefore this
discrepancy is probably not significant.

We find that whilst the normalisation determined from the optical
constraints reproduces the counts well at intermediate luminosities,
our models consistently overpredict the number of faint sources at
$z=4$, and to a lesser extent at $z=2$, when compared to the type-1
AGN luminosity function of \citet{hasinger2005}. This inconsistency
may be partially attributable to the restriction of the
\citet{hasinger2005} to type-1 (unobscured) AGN and incompleteness of 
the optical identification. If we take the
data at face value, then given that the merger-rate of galaxies
steepens significantly toward higher redshifts, then the
\citet{hasinger2005} 
data suggests that
for a merger-driven model of quasar activity $L_{\rm peak,min}$ must
be increased by approximately an order of magnitude to
$10^{12}$~L$_{\sun}$. Such a cut-off brings the rapid fading model
into agreement with the hard and soft X-ray data at $z = 2,4$, and
eases, though does not eradicate, the excess of faint objects
predicted by the slow fading model. This modification would not alter
the agreement of either model with the optical constraints.

Invoking a cut-off at $L_{\rm peak,min} = 10^{12}$~L$_{\sun}$
corresponds to  a minimum mass of active black holes of  $\sim 3 \times
10^7$~M$_{\sun}$ in our model. Whilst the mass function of local BHs may
turn over at around this value \citep[e.g.][]{greene2007,greene2007a},
this trend is highly uncertain and direct measurements of BH masses at
higher redshift are not yet available.

However, it appears possible that the current surveys have
underestimated the space density of faint sources at $z=4$
\citep{aird2008}. Note that if the measurements of a small space
density of faint hard and soft X-ray selected quasars at $z = 4$ and
the sudden drop in the density of X-ray quasars between $z = 2$ and
$4$ consolidate, then the density of high redshift X-ray quasars, and
therefore the number of sources detectable by the next generation of
X-ray satellites, may be lower than we are predicting with our
fiducial merger-driven model. In the discussion that follows we will also
discuss results for a slow and rapid fading models with $L_{\rm
peak,min} = 10^{12}$~L$_{\sun}$.

\subsection{Extrapolating the evolution of the luminosity function
 in the merger-driven model to very high redshift}

A prominent feature in our models of the high redshift hard and soft
X-ray QLFs is the break at low luminosities in the rapid fading model
due to the cut-off $L_{\rm peak,min} = 10^{11}$~L$_{\odot}$. No such
break is observed for the slow fading model where the sources with
$L_{\rm peak} = L_{\rm peak,min}$ fade gradually. Above the break, the
X-ray QLF exhibits little dependence on the fading law.  This is
because at high redshift the dependence of the merger rate on
luminosity in our model results in a luminosity function with steeper
slope than the fading rate of sources for either fading model.
There exists therefore a degeneracy in the prediction of the number of
faint sources between the slope of the assumed fading law and
the minimum peak luminosity.  Values of $L_{\rm peak,min} <
10^{11}$~L$_{\odot}$ are certainly plausible at high redshift;
theoretical models for the formation of the first BHs suggest
that seed black holes with masses between $100$ and $10^5$~M$_{\sun}$
may form as early as $z = 20$ \citep[e.g.][]{madau2001,
volonteri2006a}.  In Section~6 we explore the possibility of a lower
value of $L_{\rm peak,min} = 10^9$~L$_{\sun}$ ($M_{\rm bh,min} \approx
3 \times 10^{4}$~M$_{\sun}$).

\subsection{Alternate models for the early growth of supermassive black holes} 

The mechanism for the growth of very high redshift BHs is very
uncertain. Estimates for the formation of the first, or seed, BHs vary
widely.  BHs with masses $10^2 - 10^3$~M$_{\sun}$ may form from
population-III remnants at $z=20$ or earlier \citep{madau2001}.  To
grow BHs as large as a few $\times 10^9$~M$_{\sun}$ by $z \sim 6$ from
a pop-III remnant mass seed requires more or less continuous accretion
if the accretion rate Eddington limited
\citep[e.g.][]{archibald2002}. \citet{li_hernquist2007} find that in a
merger-driven scenario, BHs large enough to power the brightest $z
\sim 6$ quasars may be grown from seed BHs with mass
$10^5$~M$_{\sun}$. In this model most progenitors in the
(hierarchical) merging history of the SMBH grow at sub-Eddington rates
for the majority of the time. The overall growth, however, is
dominated by the fastest growing BHs at each step in the hierarchy. In
aggregate the growth is similar to almost continuous Eddington limited
accretion of a single seed BH.  Following \citet{rhook2006} we
consider a passive evolution scenario in which a fixed comoving
density of BHs evolve by accreting at their Eddington limit with a
fixed duty cycle $f_{\rm duty}$. In this model for the early BH
growth, the ensemble average luminosity of a BH increases with
decreasing redshift as,

\begin{eqnarray}
  L &=& L(z=6) e^{-(t_{z=6}-t_z)f_{\rm duty}/\kappa},\\
  \kappa &=& \frac{c \sigma_e}{4\pi G m_{\rm p}}\frac{\epsilon_{\rm
      acc}}{1-\epsilon_{\rm acc}},\\ \nonumber &\simeq& 5 \times
  10^8~{\rm yrs}\frac{\epsilon_{\rm acc}}{1-\epsilon_{\rm acc}},
\end{eqnarray}

\noindent where $\sigma_e$ is the Thompson scattering cross section.

The exponential growth of BHs in this model makes it unsuitable for
modelling the QLF below  $z \sim 6$. We therefore consider this model
as a possible alternative to the merger-driven evolution only at $z>6$.
We adopt the rapid fading merger-driven model with
$L_{\rm peak,min} = 10^{9}$~L$_{\sun}$ at $z < 6$. 

Extending the absorption laws used in the previous section, which were
determined from estimates of the average column density along the line
of sight to a pair of merging galaxies, is probably not meaningful for
continuously accreting BHs.  We therefore simply assume that a fixed,
luminosity independent, fraction of the accreting BHs are unobscured,
and that the rest are Compton-thick (completely obscured in the
optical and X-ray).  The total duty fraction is then made up of the duty
fraction of BHs that are accreting in an X-ray luminous phase $f_{\rm
 lum}$ and those that are
accreting in a Compton thick phase $f_{\rm obsc}$.

Note that for a model in which the BHs are growing continuously
(although with some duty cycle) the concept of a fading law is also  no
longer meaningful since we are assuming that there is always enough
gas for the BHs to accrete at their Eddington limit.

We have explored the evolution above $z=6$ for two values of the total
duty fraction, $f_{\rm  duty} = 0.1, 1.0$. The fraction of obscured
sources at high redshift is unknown but is certainly significant at
lower redshift \citep[e.g.][]{risaliti1999,martinez2007}.  We assume 
that half of the sources are totally obscured in each case, which is
consistent with the constraints from population synthesis models of
the CXRB \citep{gilli2007}.

For our models with Eddington limited BH growth at $z>6$ our
assumptions governing the accretion rate and intrinsic absorption
change discontinuously at $z=6$ in such a way that the observed hard
X-ray QLF changes smoothly.  This alternate model for the early growth
of SMBHs should thus be regarded only as a demonstration of the
uncertainty in the growth rate of BHs, and therefore the number of
detectable X-ray quasars, at very high redshifts.

In Section~\ref{future} we explore the predictions for the number of
quasars detectable at X-ray wavelengths for both passive and
merger-driven BH growth scenarios. We first compare our merger-driven
models to the constraints from  the observed flux distribution of
X-ray sources and the CXRB.

\section{X-ray number counts and redshift distribution at faint flux
  levels and the early growth of supermassive black holes} \label{predictions}

\subsection{X-ray source counts and redshift distributions}

\begin{figure*}\label{plot5}
  \vspace*{120mm} \includegraphics{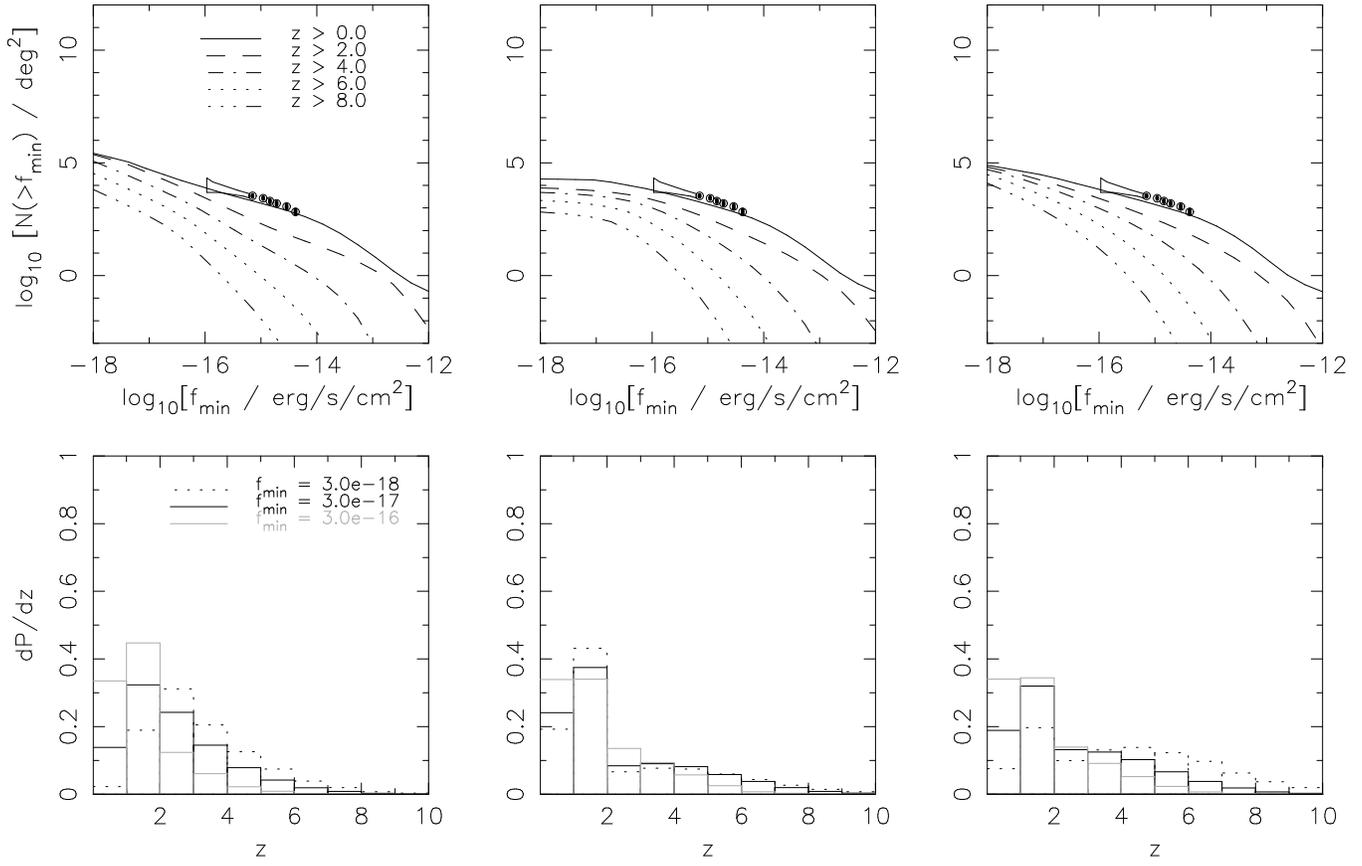}
\caption{Results for the expected number counts of sources in the
  $2-10$~keV band for the model with slow fading (left) and rapid
  fading with $L_{\rm peak,min} = 10^{11}$~L$_{\odot}$ (middle) and
  $L_{\rm peak,min} = 10^{9}$~L$_{\odot}$ (right).  Note that the
  models have been supplemented with observational QLFs below $z = 2$
  as described in the text. The top panel shows the counts integrated
  from $z = 0,2,4,6,8,10$ to $z = 20$ as labelled. The hollow markers
  with error bars (bow-tie) are the estimate of the resolved
  (unresolved) source density in the Chandra deep field by Miyaji \&
  Griffiths~(2002). The bottom panel shows the redshift distribution
  of the counts for sources above flux levels of $3.e-18$
  (\emph{dotted} line), $3.e-17$ (\emph{solid} line) and $3.e-16$
  (\emph{grey} line) in cgs units.}
\end{figure*}

\begin{figure*}\label{plot6}
  \vspace*{65mm} \includegraphics{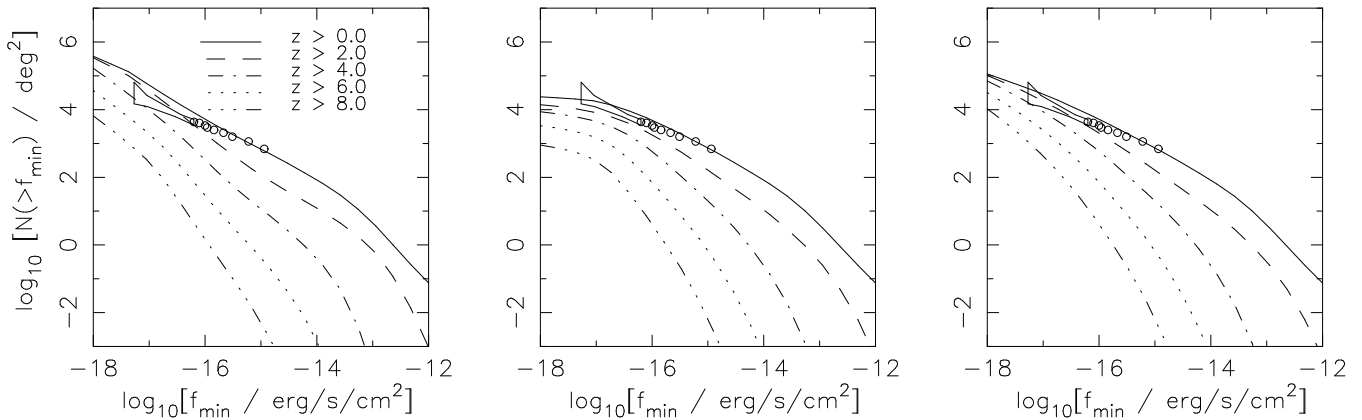}
\caption{Results for the expected number counts in the $0.5-2$~keV
band for the models in Figure~5. At $z<2$ the models have been
supplemented with the LDDE2 fit to the measured $0.5-2$~keV QLF from
Miyaji, Hasinger \& Schmidt~(2000). The hollow markers with error bars
(bow-tie) are the estimate of the resolved (unresolved) source density
in the Chandra deep field by Miyaji \& Griffiths~(2002).}
\end{figure*}

The main aim of our modelling is to predict the X-ray source counts
and the corresponding redshift distribution at faint flux levels.

The X-ray source counts (``the ${\rm log}N-{\rm log}S$ relationship'')
are well described by a double power-law, with the slope flattening
for fluxes below $\sim 1.5 \times 10^{14}$~erg~s$^{-1}$ in the soft
X-ray band and below $\sim 4.5\times10^{15}$~erg~s$^{-1}$ in the hard
X-ray band \citep[e.g.][]{cowie2002,moretti2003}. More recently, the
constraints on the number counts have been extended to lower
sensitivities by fluctuation analyses of the unresolved
background. These constraints generally still allow for an upturn in
the ${\rm log}N-{\rm log}S$ relationship below the detection limit 
for resolved sources \citep[e.g.][]{hickox2006}.

In the top panel of Figure~5 we plot the source counts as a function
of limiting flux sensitivity in the hard X-ray ($2-10$~keV) band and
in Figure~6 for the soft ($0.5-2$~keV) band.  The \emph{left} panels
show the results for the slow fading model, the \emph{middle} for the
rapid fading model with $L_{\rm peak,min} = 10^{11}$~L$_{\sun}$ and
the \emph{right} for the rapid fading with $L_{\rm peak,min} =
10^{9}$~L$_{\sun}$. In each case the over-plotted data is taken from
the fluctuation analysis of \citet{miyaji2002} for the Chandra Deep
Field North (CDF-N). Recall that below $z=2$ we have supplemented our
model for the X-ray QLF with a fit to the observed QLFs (calculated at
discrete redshifts $z = 0.01, 0.6, 1.2$). We have used the PLE model
from \citet{barger2005} for the hard X-ray band and the LDDE2 model
from \citep{miyaji2001} for the soft X-ray band. Therefore by
construction our models should reproduce the observed source counts if
our model is a reasonable approximation at $z \geq 2$.

Our models are consistent with the constraints on the hard X-ray
counts for both the rapid and slow fading laws. However, for the soft
X-ray counts only the rapid fading models are consistent with the
observed counts. Using the slow fading law our model slightly
overshoots the faint soft X-ray counts and lies above the limit
suggest by the fluctuation analysis of \citet{miyaji2002}, as is
expected from the inconsistency with the measured QLF at $z \sim 2$
discussed in Section~4.1. For the slow and rapid fading models with
$L_{\rm peak,min} = 10^{12}$~L$_{\sun}$, the number counts below the
current detection levels are reduced, bringing the slow fading model
into agreement with the fluctuation analysis and resulting in a flat
source distribution at low flux levels for the rapid fading model.

In the lower panel of Figure~5 we plot the normalised redshift
distribution for the source counts above three flux/sensitivity
levels. The \emph{grey} line is for the approximate sensitivity of
current X-ray satellites ($3\times 10^{-16}$~ergs/s/cm$^2$), the
\emph{solid black} line for an order of magnitude fainter ($3\times
10^{-17}$~ergs/s/cm$^2$) and the dotted line for the goal sensitivity
of a 1~Ms observation with XEUS ($3\times 10^{-18}$~ergs/s/cm$^2$, but
see the discussion in Section~\ref{future}). We note the strong
dependence of the redshift distribution on the fading law. For the
slow  fading model the probability for source detection peaks
at $z \sim 1-3$ for each of the chosen flux limits, with the peak
shifting to slightly higher redshift for the lower flux limits. For
this model the source counts remain dominated by fading objects at
intermediate redshifts where the space  density of quasars is at
its peak.

For the rapid fading models the source distribution is sharply peaked
at $z \sim 1$ for the brightest flux limit. This is not surprising:
this is where most of the known sources lie and we have used the
observed data at low redshifts as our model QLF. At fainter flux
limits a second broader peak around $z \sim 4$ emerges for the rapid
fading models.  The broad high redshift peak at faint flux level in
our model is due to the steepening of the merger rate at $z >2$
combined with the increase in the comoving volume element out to $z
\sim 3.5$.

The emergence of two distinct populations at faint flux levels, at low
and high redshift, can be explained physically by an evolution of the
Eddington ratio and the characteristic lifetime with
redshift. Observationally, optically selected high redshift and/or
high luminosity AGN display near Eddington accretion rates \citep[see,
e.g.][]{mclure2004, kollmeier2006}, whereas hard X-ray selected
sources at $z<1$ appear to have much lower accretion rates
\citep[e.g.][]{babic2007}. At decreasing flux levels surveys will become
sensitive to both the very faint sources accreting at low redshift and
sources accreting at higher rates at high redshift, and therefore the
source distribution should become double peaked.

The predicted high redshift peak for the rapid fading model is, as
expected, more prominent for the scenario where more abundant, smaller
mass BHs may power quasars.  For the case with $L_{\rm peak,min} =
10^9$~L$_{\sun}$ almost a quarter of sources are predicted to be above
$z \sim 6$ at the lowest flux level plotted.  The fraction of sources
expected above $z = 4,6,8$ for each of the models at the three flux
limits shown in Figure~5 is tabulated in Table~1.

\begin{table}
  \begin{tabular}{|l|l|l|l}
    \hline
$f_{\rm min}$~(erg/s/cm$^2$)&Fading law&$z_{\rm min}$&$x_{\rm obs}$\\
\hline
$3.e-16$ & Slow & ~4 & $3.1\times 10^{-2}$\\
& & ~6 & $1.9 \times 10^{-3}$\\
& & ~8 & $5.9 \times 10^{-5}$\\
& Rapid & ~4 & $6.0 \times 10^{-2}$\\
& $L_{\rm peak,min} = 10^{11}$~L$_{\sun}$& ~6 & $4.2 \times 10^{-3}$\\
& & ~8 & $1.3 \times 10^{-4}$\\
& Rapid & ~4 & $6.0\times 10^{-2}$\\
& $L_{\rm peak,min} = 10^{9}$~L$_{\sun}$& ~6 & $4.2 \times 10^{-3}$\\
& & ~8 & $1.3 \times 10^{-4}$\\
\hline
$3.e-17$ & Slow & ~4 & $1.3\times 10^{-1}$\\
& & ~6 & $1.8 \times 10^{-2}$\\
& & ~8 & $1.6 \times 10^{-3}$\\
& Rapid & ~4 & $2.1 \times 10^{-1}$\\
& $L_{\rm peak,min} = 10^{11}$~L$_{\sun}$& ~6 & $4.8 \times 10^{-2}$\\
& & ~8 & $5.5 \times 10^{-3}$\\
& Rapid & ~4 & $2.0\times 10^{-1}$\\
& $L_{\rm peak,min} = 10^{9}$~L$_{\sun}$& ~6 & $4.0 \times 10^{-2}$\\
& & ~8 & $4.5 \times 10^{-3}$\\
\hline
$3.e-18$ & Slow & ~4 & $2.3\times 10^{-1}$\\
& & ~6 & $4.8 \times 10^{-2}$\\
& & ~8 & $8.2 \times 10^{-3}$\\
& Rapid & ~4 & $3.2 \times 10^{-1}$\\
& $L_{\rm peak,min} = 10^{11}$~L$_{\sun}$& ~6 & $1.2 \times 10^{-1}$\\
& & ~8 & $3.0 \times 10^{-2}$\\
& Rapid & ~4 & $4.6\times 10^{-1}$\\
& $L_{\rm peak,min} = 10^{9}$~L$_{\sun}$& ~6 & $1.7 \times 10^{-1}$\\
& & ~8 & $4.3 \times 10^{-2}$\\\hline
\end{tabular}
\caption{Table of the fraction of sources with observed 
$0.5-2$~keV fluxes above $f_{\rm min}$ and redshift above $z_{\rm min}$
for the 3 merger-driven models considered.}
\end{table}

\begin{figure*}\label{plot7}
  \vspace*{75mm} \includegraphics{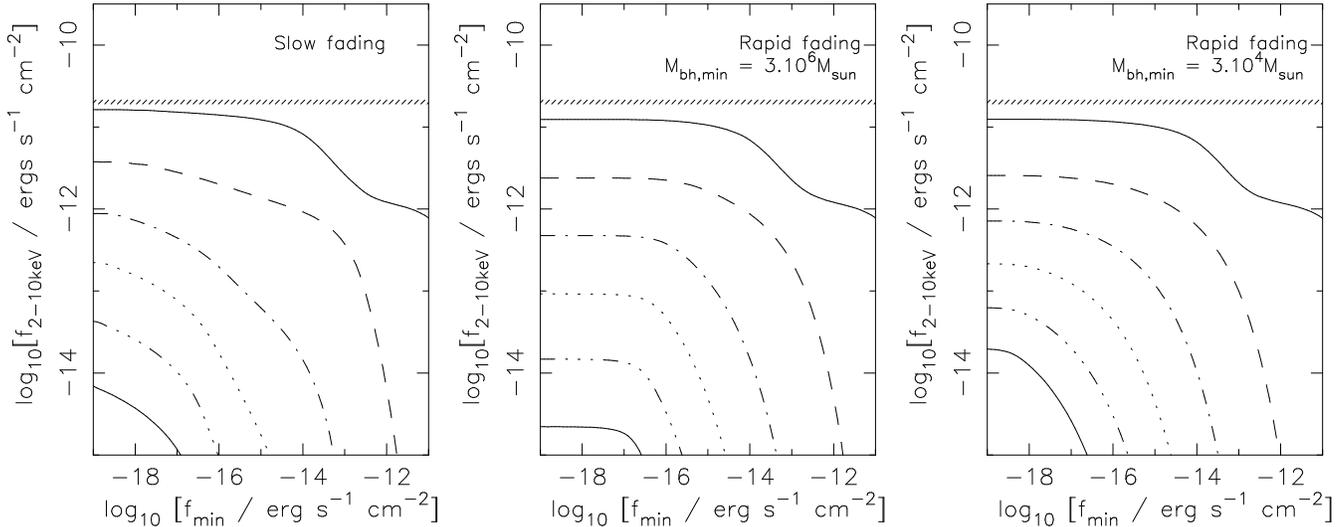}
\caption{Integrated hard ($2-10$~keV) X-ray flux due to AGN for the models in
  Figure~5.  The hatched region shows the estimate of the total
  extragalactic CXRB in the $2-10$~keV band from
  \citet{moretti2003}.The \emph{left} panel is for slow 
  fading with $L_{\rm peak,min} = 10^{11}$~L$_{\odot}$. The
  \emph{middle} (\emph{right}) panels show the results for rapid
  fading with $L_{\rm peak,min} = 10^{11}$~L$_{\odot}$ ($L_{\rm
  peak,min} = 10^{9}$~L$_{\odot}$)}
\end{figure*}

\begin{figure*}\label{plot8}
  \vspace*{75mm} \includegraphics{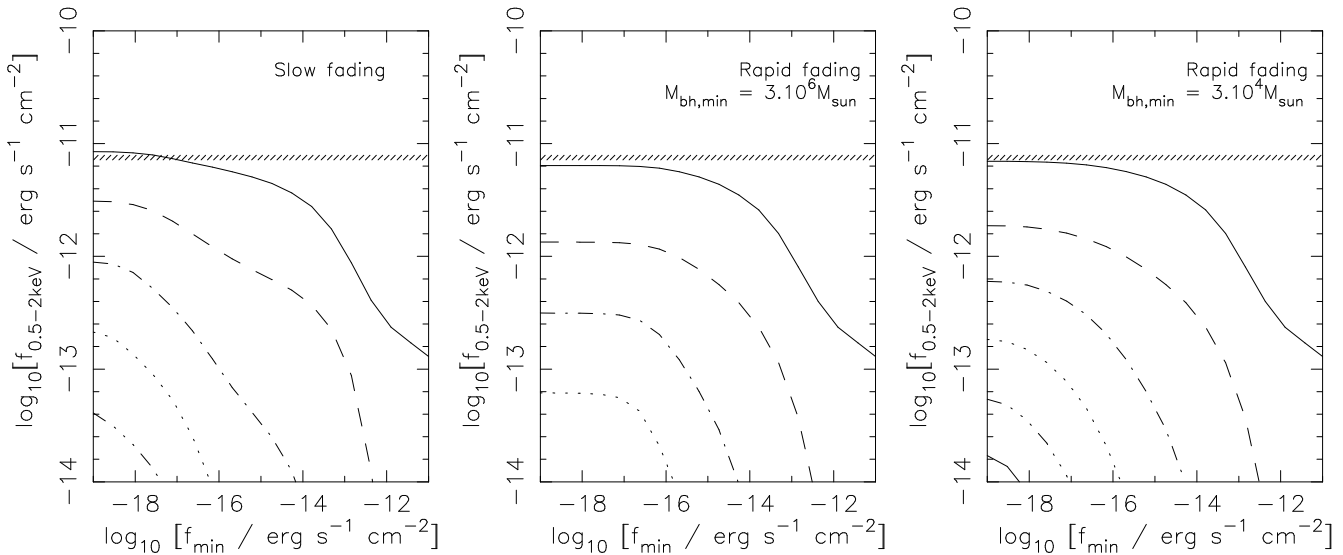}
\caption{Integrated soft ($0.5-2$~keV) X-ray  flux due to AGN for the models in
Figure~6. The hatched region shows the estimate of the total
extragalactic component of the CXRB in the $0.5-2$~keV band from
\citet{moretti2003}.}
\end{figure*}

\subsection{The integrated  X-ray background}\label{XRB}

\begin{figure*}
  \vspace*{60mm} \includegraphics{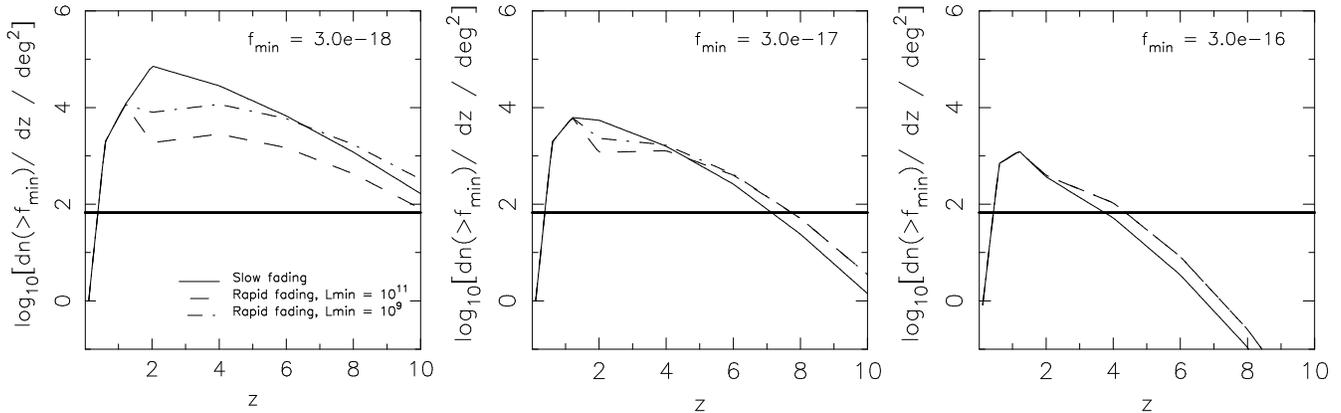}
\caption{Redshift distribution of predicted counts in the $0.5-2$~keV band for each model; slow fading (solid line), rapid fading with $L_{\rm peak,min} = 10^{11}$~L$_{\sun}$ (dashed line) and $L_{\rm peak,min} = 10^{9}$~L$_{\sun}$ (dot-dashed line) for three flux limits as labelled in cgs units. The horizontal line corresponds to a density of one source per XEUS WFI FOV.}
\end{figure*}

\begin{figure*}\label{plot10}
  \vspace*{60mm} \includegraphics{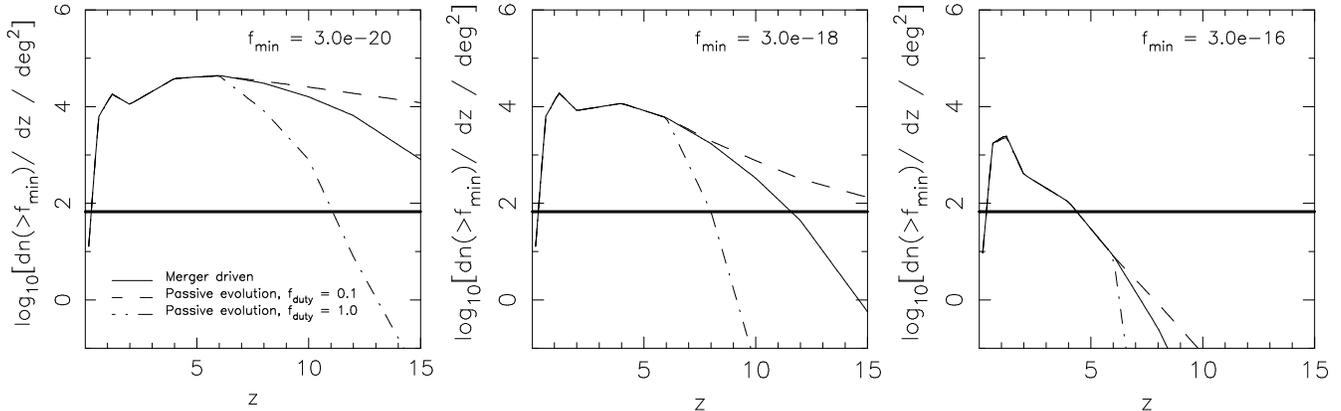}
\caption{Redshift distribution of counts in the $0.5-2$~keV band for the
  rapid-fading merger driven model and passive evolution models as
  described in text; Merger-driven (solid line), passive evolution
  with total duty cycle $f_{\rm duty} = 0.1$ (dashed line) and $f_{\rm
  duty} = 1.0$ (dot-dashed line) for three flux limits as labelled in
  cgs units. Note that the lowest flux limit is the anticipated
  sensitivity for the Generation-X mission, the middle for a 1 Ms
  observation with XEUS, and the right for a point-source within reach 
  of current instruments.The horizontal line corresponds
  to a density of one source per XEUS WFI FOV (7~arcmin)$^2$, which
  is similar in area to the proposed FOV for Generation-X.}
\end{figure*}

The Cosmic X-ray Background (CXRB) and its unresolved component
provides an important consistency check for models of the number of
faint sources.

We may calculate the contribution of quasars in a redshift band
$(z,z+dz)$ to the X-ray background in an observed band $X$ by
integrating over the QLF,
\begin{equation}\label{faint_flux}
\frac{df_X}{dz}dz = \int_{L_X = f_{\rm min}4\pi D_L^2(z)} \left.\frac{dn}{dlogL_X}\right|_{\rm obs}(z)\frac{dV_c}{d\Omega}\frac{L_X}{[D_L(z)]^2}dlogL_X,
\end{equation}

\noindent where $D_L$ is the luminosity distance and the limits of
integration are determined from the limiting sensitivity $f_{\rm
min}$.

In Figures~7 and 8 we compare the total CXRB flux predicted for the
slow fading model, and the rapid fading models with low and high
$L_{\rm peak,min}$, to the measurement of the total CXRB from
\citet{moretti2003}. This contribution is plotted as a function of the
minimum detectable flux for sources with redshifts above $z =
0.01,2,4,,6,8$ and $10$. We note that the resolved fraction is
sensitive to uncertainties in the absolute value of the CXRB. We
restrict our analysis of the resolved fraction to energies below
10~keV, where variations in the normalisation between experiments
differ by $\sim 10$~per~cent \citep[see, e.g.][]{moretti2003}.

In the $2-10$~keV band, we find that all models are consistent with
the total measured CXRB. Our models predict that $\sim 70
-78$~per~cent of the total $2-10$~keV CXRB is due to AGN with fluxes
above the best current sensitivity level in this band ($\sim 1.4\times
10^{-16}$~cgs). This result is consistent with the $\sim 80$~per~cent
resolved fraction measured by \citet{worsley2005} and
\citet{hickox2006}.  Our models remain consistent with the total CXRB
at lower flux limits. At a sensitivity of $3 \times 10^{-18}$~cgs - a
sensitivity within reach of next generation instruments (see
Section~5.3) - the resolved fraction increases to $\sim 87$~per~cent
depending on the model.  Our models thus still leave room for the very
hard spectrum sources need to make up the CXRB at energies $> 8$~keV
\citep{worsley2005} and/or the population of star-forming galaxies
expected to contribute significantly at low flux levels
\citep[e.g.][]{bauer2004}. We note that our calculated resolved fractions
also depend on the analytic fits we have used below $z = 2$. Using
the more recent LDDE model from \citet{silverman2007} we obtain 
lower resolved fractions ($\sim 62 -75$~per~cent due to sources above
the current detection level), consistent with the more stringent
optical selection criteria for this sample.

In the soft X-ray band ($0.5 - 2$~keV), the fraction of the total CXRB
flux due to sources above the current sensitivity level ($\sim 2.5
\times 10^{-17}$~cgs) in our models is again consistent with the $80 -
90$~per~cent found by \citet{worsley2005} and \citet{hickox2006}.
However, at fainter flux levels the slow fading model overpredicts the
total CXRB, again reflecting the excess of faint sources at $z \sim 2$
compared to the observed QLF. This contribution again depends
on the faint end behaviour of the fits to the data that we have
adopted below $z \sim 2$. As mentioned, for the soft X-ray band we
have chosen to use the LDDE2 model from \citet{miyaji2001} which is
constructed to reproduce $\sim 90$ of the total soft X-ray background
when integrated out to $z \sim 5$. Alternate faint end extrapolations
may somewhat ease this excess. Using the \citet{hasinger2005}
LDDE fit to the LF of type-1 AGN below $z=2$ we find that the slow
fading model saturates, but does not overpredict, the soft X-ray
background. However the \citet{hasinger2005} LDDE fit to the soft
X-ray type-1 QLF only accounts for $\sim 35$~per~cent of the soft band
CXRB, reflecting the fact that the sample of type-1 AGN used in
\citet{hasinger2005} account for only around 30~per~cent of sources at
faint and bright flux levels.

The resolved fractions are naturally lower for the slow and rapid
fading models with $L_{\rm peak,min} = 10^{12}$~L$_{\sun}$ (chosen to
reproduce the soft X-ray QLF at $z \sim 2$), and with this choice the
slow fading model remains consistent with the total CXRB.

Recently, \citep{worsley2005} and \citet{hickox2006} have pushed the
limit for the unresolved background further by taking into account the
stacked emission from galaxies detected with {\it HST} and {\it
IRAC}. Indeed we find that if we integrate the flux due to sources
{\it below} the current detection level and compare this to the
measurement of the unresolved component as derived by
\citet{hickox2006}, the soft and hard X-ray components are both too
large for the slow fading model. Models like this for which sources
with relatively flat spectra recover the entire unresolved CXRB in the
soft and hard bands are likely to be in conflict with the overall
measured shape of the CXRB above $8$~keV
\citep[e.g.][]{worsley2005,comastri1995, gilli2007}. This consistency
check argues further against the slow fading model as we have applied
it, but we note that a flattening of the fading law slope for low mass
black holes would alleviate the inconsistencies with the CXRB and the
soft X-ray source counts. In particular the slow fading model
with $L_{\rm peak,min} = 10^{12}$~L$_{\sun}$ saturates, but does not
over-predict, the unresolved component of the soft CXRB.

In a recent paper, \citet{salvaterra2007} predict the contribution of
high redshift AGN to the unresolved CXRB for a merger tree based model
for the Eddington limited growth of BH seeds. They find that $\sim
5$~per~cent of the unresolved $2-10$~keV CXRB will be due to sources
at $z>6$.  This is similar to the $\sim 10$~per~cent that we predict
for the rapid fading model with $M_{\rm bh,min} = 3 \times
10^{4}$~M$_{\sun}$, but significantly larger than we would predict for
models with larger $M_{\rm bh,min}$.

In the next section we focus on the predictions for the rapid fading
models, since these are consistent with all available data, but have
included the predictions for the slow  fading model for
reference.

\subsection{Future hard X-ray surveys and the early growth of
 supermassive black holes}\label{future}

The sensitivities of the deepest X-ray surveys to date are photon
limited and therefore independent of the background intensity.  This
may not be the case for the effective collecting area(s) and angular
resolution(s) anticipated for XEUS and Constellation-X. Estimates of
the point-source sensitivity for future instruments therefore
potentially become dependent on the assumed unresolved extragalactic
component (due to quasars and star-forming galaxies) and its flux
distribution, as well as the effective collecting area and photon
extraction radius (or resolution). Due to uncertainties in the
decomposition of the X-ray background into contributions from the
galaxy, star-forming galaxies and quasars, it is somewhat uncertain
what the confusion limit of these telescopes will be.  
The projected point-source sensitivities for the next generation
telescope peak for soft X-ray energies. At these energies the soft
thermal galactic component is a significant portion of the background
\citep[see, e.g.][]{parmar1999}, however the contribution of
unresolved point sources may also play an important role if the
ambitious goals for the resolution are not achieved. Since our models
predict the contribution of quasars to this background, the expected
sensitivity of an instrument like XEUS to high redshift sources may
depend on the modelling of faint sources.

\citet{hasinger2006} discuss the anticipated point source sensitivity
of Constellation-X and XEUS. Assuming that the observations are
confusion limited when there are fewer than 40 ``beams'' per source,
combined with estimates of the background due to unresolved
extragalactic sources, galactic emission and cosmic-rays,
\citet{hasinger2006} project that a 1~Ms observation with XEUS will
yield a point source sensitivity of $3 \times
10^{-18}$~erg~s$^{-1}$~cm$^{-2}$, approximately $200$ times better
than that of XMM-Newton. This estimate presumes that XEUS's goal
resolution of 2~arcseconds will be achieved. Achieving this resolution
is expected to be extremely challenging, and the required resolution
of 5~arcseconds is perhaps more realistic \citep{hasinger2006}. For this
resolution the sensitivity degrades to $\sim 2 \times
10^{-17}$~erg~s$^{-1}$~cm$^{-2}$ for a $1$~Ms integration.  Similarly,
for Constellation-X \citet{hasinger2006} project a point source
sensitivity of $2 \times 10^{-17}$~erg~s$^{-1}$~cm$^{-2}$ for a 1~Ms
observation for the Constellation-X goal resolution of 5~arcseconds.

Assuming a circular beam and applying the above confusion criterion
suggests that XEUS will be able to identify point sources with
densities below $\sim 8.6 \times 10^{4}$~deg$^{-2}$ if it reaches its
goal resolution. Comparison with the source densities in our models
suggests that XEUS should not suffer from confusion due to faint AGN
at this flux limit for the rapid fading models, but would do so below
$\sim 10^{-17}$~erg~s$^{-1}$~cm$^{-2}$ for the slow fading model.  The
reduced sensitivity is not merely attributable to the excess in the
predicted number of faint sources at $z \sim 2$, since the density of
sources above $z \sim 4$ alone are sufficient to stop XEUS from
reaching its projected sensitivity in this model.  This suggests that
the confusion limit for XEUS may be somewhat sensitive to the way
quasars fade at $2<z<6$.  However, we note that the same slow fading
model with a higher minimum peak luminosity $L_{\rm peak,min} =
10^{12}$~L$_{\sun}$ would not predict a degradation of the estimated
point source sensitivity due to source confusion by faint
AGN. Obviously the situation would be less favourable if the
resolution goal of $2$~arcseconds HEW could not be achieved. If the
resolution is reduced to the required value of 5~arcminutes, we find
the density of sources above $\sim 5 \times
10^{17}$~erg~s$^{-1}$~cm$^{-2}$ exceeds the density defined by the
confusion limit ($\sim 1.4 \times 10^{4}$~deg$^{-2}$) for the rapid
fading model with $L_{\rm peak,min} = 10^9$~L${\sun}$, and even the
rather flat low-flux source density in the rapid fading model with
$L_{\rm peak,min} = 10^{11}$ would limit the sensitivity to sources
with fluxes above above $\sim 3 \times
10^{17}$~erg~s$^{-1}$~cm$^{-2}$. Similar sensitivities would be
applicable to Constellation-X if it reaches its goal resolution.

The planned
WFI\footnote{ftp://ftp.xray.mpe.mpg.de/people/bol/xeus/XEUS\_150108.pdf}
is designed to have a field of view (FOV) of (7~arcminutes)$^2$
[compared to (5~arcminutes)$^2$ for Constellation-X,
\citet{garcia2007}].  An unclustered distribution of sources with
density $\sim 6.7 \times 10^{1}$~deg$^{-2}$ will then contain one
source per WFI FOV. We highlight this density as a horizontal line in
Figures 9 and 10 for reference.  For the rapid-fading models, for
which observations should not be confusion limited down to the goal
sensitivity of XEUS, we predict significant numbers of sources (more
than one per WFI FOV) out to $z \sim 10$.  To put these sensitivities
in context for models of SMBH growth, in Figure~11 we plot the
$0.5-2$~keV flux in the observed frame for BHs accreting at their
Eddington limit (using the \citet{hopkins2006c} $z$-independent SED)
as a function of the source redshift. The lowest mass BH that may be
seen accreting at its Eddington limit at $z \sim 6 - 10$ for a $1$~Ms
observation with XEUS is in the range $\sim 10^{5}$ - $5 \times
10^{5}$~M$_{\sun}$.  For a model in which only BHs larger than $\sim 3
\times 10^{7}$~M$_{\sun}$ are active (\emph{i.e.} $L_{\rm peak,min} =
10^{12}$~L$_{\rm sun}$), the redshift limit up to which we predict
more than one source per XEUS field is reduced to 8 (7) for the slow
(rapid) fading model.

Contrastingly a telescope with 5~arcsecond resolution, which as
mentioned above may be confusion limited below $\sim 3 \times
10^{-17}$~erg~s$^{-1}$~cm$^{-2}$, may be sensitive enough to detect to
more than one source per XEUS (Constellation-X) FOV up to $z \sim 7$
($z \sim 6$), but detection of significant numbers of sources at
higher redshifts would require surveying many XEUS/Constellation-X
fields (see Figure~9). This result changes little for the case $L_{\rm
peak,min} = 10^{12}$~L$_{\odot}$, with approximately one source per
XEUS field out to $z \sim 6$ predicted independently of the fading
law.

The confusion limits we have discussed represent the best case for
each model given the assumptions in \citet{hasinger2006} for the
contribution from galactic emission and star-forming galaxies, and the
true confusion limits may therefore be worse if these have been
underestimated.

Improvements in the resolution of hard X-ray telescopes will naturally
improve the confusion limit for the point source sensitivity (where it
is limited by discrete sources), particularly for the case of rapid
fading in which the source counts rise slowly with decreasing
flux. NASA's mission concept Generation-X has a goal resolution of
$0.1$~arcseconds and $100$~m$^2$ of collecting area yielding a
photometric sensitivity of $\sim 2.2 \times
10^{-20}$~ergs~s$^{-1}$~cm$^{-2}$ at $0.1 -
10$~keV\footnote{http://www.psu.edu/dept/csrp/missions\_awarded\_genx.htm}.
This resolution would certainly avoid confusion due to faint quasars,
even for the slow fading model where there are many faint foreground
sources confusing the detection of sources at $z \gtrsim 5$.

With the proposed sensitivity of Generation-X, unobscured BHs more
massive than around a ${\rm few} \times 1000~{\rm M}_{\sun}$ would be
detectable out to redshifts approaching 15 (see Figure~11). Detection
of the seeds of SMBHs is one of the prime science goals of
Generation-X.  The number of detectable objects at $z \sim 6 - 15$
will obviously depend on the masses of seed BHs and the growth history
of BHs at very high redshifts.

In Figure~10 we compare the predicted number of detectable sources for
the rapid fading merger-driven model (\emph{solid} line) to the
passive evolution models with $f_{\rm duty} = 0.1$ (\emph{dotted}
line) and $f_{\rm duty} = 1.0$ (\emph{dot-dashed} line) for three
values of the point-source sensitivity. The left panel shows the
evolution in the density of sources with fluxes above the sensitivity
goal of Generation-X. The space density of observable sources at $z
\sim 15$ depends very sensitively on the model assumed. The FOV of
Generation-X is expected to be at least (5~arcminutes)$^2$, with
fields as big as (15~arcminutes)$^2$ considered in the design study
\citep{windhorst2006}. Here we assume the FOV of Generation-X is the
same as for XEUS for ease of comparison.  For the merger driven model,
with a minimum active BH mass of $\sim 3 \times 10^{4}$~M$_{\sun}$ we
predict significant numbers of sources, still around 10 per FOV, out
to $z = 15$. The situation is even more promising if the BHs are
evolving passively with a small duty cycle, with up to $\sim 100$
sources per field up to $z = 15$. If the BHs are evolving rapidly
there is still more than one source per field out to $z \sim 11$, but
the density drops very rapidly with increasing redshift for a scenario
in which all BHs are growing via accretion at their Eddington
limit. At the goal sensitivity of XEUS, the maximum redshift at which
there is one source predicted per FOV also varies widely; ranging from
$z\sim 8$ for continuous growth to $z \sim 15$ for slowly growing BHs.

\section{Summary and conclusions}

We have presented a hybrid model for the redshift evolution of the
X-ray emission connected with the fuelling of supermassive black holes
with the aim of assessing the prospects of detecting quasars via their
X-ray emission at $z>6$. At $z \lesssim 2$ we have used the observed
X-ray luminosity functions, at $2<z<6$ we have adopted a CDM
merger-driven model for the evolution of the emission from
supermassive black holes combined with a slow and a rapid fading law
and we have explored a range of assumptions for the growth of
supermassive black holes at $z>6$.

For appropriate choices for the efficiency of black hole formation in
a dark matter halo and the characteristic quasar lifetime, our model
is in good agreement with the observed optical and soft and hard X-ray
quasar luminosity functions at $0<z<6$ for both the slow and the rapid
fading law. The only disagreement that occurs is for the faint end of
the X-ray luminosity functions where our merger model combined with
the slow fading model suggested by \citet{hopkins2005b} based on
detailed numerical simulations predicts too many faint objects. It
consequently also overpredicts the soft X-ray background suggesting
that the slow fading model of \citet{hopkins2005b} cannot extend to
black hole masses much below $10^{7}$~M$_{\sun}$.  With a more rapid
fading law our models are well within the limits of the integrated
CXRB, leaving room for the population of star-forming galaxies
expected at low flux densities, and the population of Compton-thick
sources needed to explain the overall spectral shape of the CXRB.
 
The main aim of our study is to assess the prospects of planned and
proposed X-ray missions like Constellation-X, XEUS and Generation-X to
study the build-up of black holes at $z>6$. With a point source
sensitivity of $3\times 10^{-17}$~ergs~s$^{-1}$~cm$^{-2}$ for a 1~Ms
exposure, Constellation-X will detect significant numbers of black
holes at $z\sim 5-6$ but will probably not yet reach the necessary
sensitivity to detect significant numbers of sources at $z>6$. This
conclusion holds independently of our assumptions for the growth of
BHs at $z>6$.  For XEUS with its anticipated ten times superior point
source sensitivity, the prospects for detecting sources at $z>6$ is
much more favourable. Up to 17 per cent of the approximately 100
sources in the 49 square arcminute FOV expected for the rapid fading
model with a minimum active black hole mass of $\sim 3 \times
10^4$~M$_{\odot}$ are expected to be at $z>6$. If our merger driven
model can be extrapolated to $z>6$ or if black holes grew with a duty
cycle of about 10~per~cent, XEUS would detect significant numbers of
black holes with a rather flat redshift distribution out to $z\sim
10$.  In either case, observable X-ray emission would have to
accompany the growth of supermassive black holes starting from seed
black holes of $\ga 10^{4}$~M$_{\sun}$ or smaller.  For the more
remote prospect of Generation-X, with an anticipated point source
sensitivity yet a factor 100 better again, the flat redshift
distribution could extent to $z\sim 15$ and beyond.

If black holes grew much faster at $z>6$ as in our Eddington limited
growth model with a duty fraction of unity, which would be required
for Eddington limited growth of the most massive black holes from
stellar mass seed black holes, neither XEUS nor Generation-X would
detect many black holes at $z \gtrsim 10$. However, while such rapid
growth may occur and may indeed be necessary for the most massive
black holes at $z \sim 6$, it appears unlikely that this will be the
norm for the majority of black holes.  The prospects that XEUS (and
Generation-X) will unravel the growth history of black holes at $z>6$
and beyond are thus excellent.

\begin{figure}
  \vspace*{110mm} \includegraphics{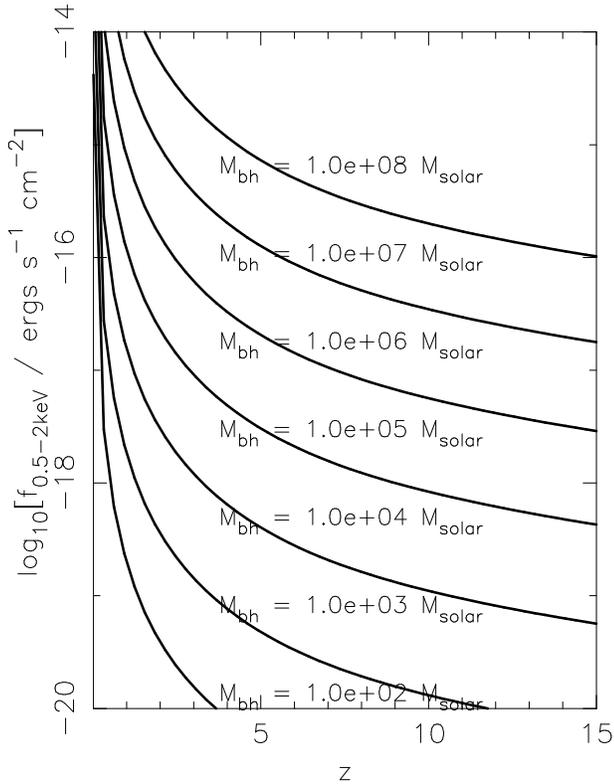}
\caption{Observed $0.5-2$~keV flux from a BH with the
\citet{hopkins2006c} SED shining at its Eddington limit as a function
of source redshift. BH masses as labelled. }
\end{figure}

\section*{Acknowledgments}
KJR was supported by an Overseas Research Scholarship, the Cambridge
Australia Trust, the School of Physics, University of Melbourne and an
\emph{EARA} visiting post-graduate scholarship hosted by MPA during the
course of this work. We would like to thank the referee for a detailed
and helpful report.

\bibliographystyle{mn2e} \bibliography{citations}

\end{document}